\documentclass[sigconf,screen,balance]{acmart}

\setcopyright{acmcopyright}
\acmPrice{15.00}
\acmDOI{10.1145/3368089.3409676}
\acmYear{2020}
\copyrightyear{2020}
\acmSubmissionID{fse20main-p102-p}
\acmISBN{978-1-4503-7043-1/20/11}
\acmConference[ESEC/FSE '20]{Proceedings of the 28th ACM Joint European Software Engineering Conference and Symposium on the Foundations of Software Engineering}{November 8--13, 2020}{Virtual Event, USA}
\acmBooktitle{Proceedings of the 28th ACM Joint European Software Engineering Conference and Symposium on the Foundations of Software Engineering (ESEC/FSE '20), November 8--13, 2020, Virtual Event, USA}

\usepackage{soul}
\usepackage{url}
\usepackage[utf8]{inputenc}
\usepackage{caption}
\usepackage{balance}

\usepackage{amsmath,amssymb,amsfonts}
\usepackage{algorithm}
\usepackage{algorithmic}
\usepackage{graphicx}
\usepackage{textcomp}

\usepackage{booktabs} 
\usepackage{siunitx}
\usepackage{amsmath,xcolor,pifont}
\usepackage{xspace,url}

\usepackage{color}
\usepackage{colortbl}
\usepackage{multirow}
\usepackage{multicol}
\usepackage{makecell}
\usepackage{tabularx}
\usepackage{listings}
\usepackage{graphicx}
\usepackage{caption}
\usepackage{subfigure}
\usepackage{enumitem}
\usepackage{microtype}
\usepackage[colorinlistoftodos]{todonotes}

\newtheorem{defn}{Definition}

\newcommand{\sys}{NNSlicer\xspace} 

\newcommand{\etal}{\textit{et al}.~}
\newcommand{\ie}{\textit{i}.\textit{e}.~}
\newcommand{\eg}{\textit{e}.\textit{g}.~}
\newcommand{\tabincell}[2]{\begin{tabular}{@{}#1@{}}#2\end{tabular}}

\begin{document}

\title{Dynamic Slicing for Deep Neural Networks}


\author{Ziqi Zhang}
\affiliation{
\institution{Key Laboratory of High-Confidence Software Technologies (MOE),
\\Dept of Computer Science, 
\\Peking University}
\city{Beijing}
\country{China}}

\email{ziqi_zhang@pku.edu.cn}
\authornote{The first two authors contributed equally. This work is partly done while Ziqi Zhang was an intern at Microsoft Research. $\dagger$ Correspondence to: Yuanchun Li, Yao Guo.}

\author{Yuanchun Li}
\affiliation{\institution{Microsoft Research}
\city{Beijing}
\country{China}}
\email{Yuanchun.Li@microsoft.com}
\authornotemark[1]
\authornotemark[2]

\author{Yao Guo}
\affiliation{
\institution{Key Laboratory of High-Confidence Software Technologies (MOE),
\\Dept of Computer Science, 
\\Peking University}
\city{Beijing}
\country{China}}

\email{yaoguo@pku.edu.cn}
\authornotemark[2]

\author{Xiangqun Chen}
\affiliation{
\institution{Key Laboratory of High-Confidence Software Technologies (MOE),
\\Dept of Computer Science, 
\\Peking University}
\city{Beijing}
\country{China}
}

\email{cherry@pku.edu.cn}

\author{Yunxin Liu}
\affiliation{\institution{Microsoft Research}
\city{Beijing}
\country{China}}
\email{Yunxin.Liu@microsoft.com}

\begin{abstract}
Program slicing has been widely applied in a variety of software engineering tasks. However, existing program slicing techniques only deal with traditional programs that are constructed with instructions and variables, rather than neural networks that are composed of neurons and synapses.
In this paper, we propose \sys, the first approach for slicing deep neural networks based on data flow analysis.
Our method understands the reaction of each neuron to an input based on the difference between its behavior activated by the input and the average behavior over the whole dataset. Then we quantify the neuron contributions to the slicing criterion by recursively backtracking from the output neurons, and calculate the slice as the neurons and the synapses with larger contributions.
We demonstrate the usefulness and effectiveness of \sys with three applications, including adversarial input detection, model pruning, and selective model protection. In all applications, \sys significantly outperforms other baselines that do not rely on data flow analysis.
\end{abstract}



\maketitle

\section{Introduction}

Program slicing~\cite{weiser1981slicing} is widely used in software engineering for various tasks such as debugging \cite{agrawal1993debugging}, testing \cite{arlt2014reducing}, and verification \cite{clarke1999program}. It aims to compute a set of statements (named \emph{program slice}) that may affect the values at some points of interest (named \emph{slicing criterion}). For example, by setting the slicing criterion to a specific output that generates an error, one can get a program slice that may be relevant to the error but much smaller in size than the whole program, thus much easier to analyze.

Existing program slicing techniques are mainly designed for traditional programs that are constructed with human-defined functions and instructions. Deep Neural Networks (DNNs), which have achieved remarkable success in many data-processing applications in recent years, can also be viewed as a special type of programs constructed with artificial neurons (a neuron is a mathematical function that receives one or more inputs and computes an output, such as the weighted sum or the maximum.) and synapses (the connections between neurons). However, the weights of synapses are learned by the machine and are usually hard for a human to understand. To the best of our knowledge, it has not been studied on whether and how a DNN can be analyzed meaningfully using program slicing techniques.

We apply the concept of program slicing to the area of DNNs and define \textbf{DNN slicing} as \emph{computing a subset of neurons and synapses that may significantly affect the values of certain interested neurons.}
Slicing a DNN is interesting for a number of reasons. First, it is a widely-concerned problem that the decisions made by DNNs are difficult to explain or debug. Program slicing, hopefully, can be used to extract the operations that lead to a decision, making it easier to interpret. Second, the size of DNN is growing rapidly in recent years, with more than 25 million parameters (180 MB in size) in a state-of-the-art computer vision model~\cite{tan2019efficientnet} and 110 million parameters (340 MB in size) in a state-of-the-art natural language understanding model~\cite{devlin2018bert}. How to improve the model efficiency has become an important research problem. To this end, we believe program slicing has the potential to help reduce the model size significantly. Last but not least, partitioning the model into important slices and less-important slices can also benefit model protection, as one can prioritize the important slices if protecting the whole model is difficult or impossible.

DNN slicing introduces several new challenges as compared with traditional program slicing. First, unlike the instructions and variables in traditional programs that are themselves meaningful, a neuron in a DNN is usually a meaningless mathematical operator, whose behavior is determined by its learned weights and connections with other neurons. Thus, it is a challenging problem to understand the behavior of each neuron based on its connections and weights.
Second, each output value of a DNN is affected by almost all neurons in the model. To generate a meaningful and concise slice, we must differentiate and characterize the neurons based on their contributions to the slicing criterion.
Finally, the data flow graphs in traditional programs are usually sparse and small-scale, while the data flow graphs of a DNN may contain millions of neurons densely connected with each other. Analyzing a graph in such a large scale poses a much higher requirement on system efficiency.

In this paper, we present \sys, a dynamic slicing technique for DNNs based on data flow analysis on neural networks.
The \textit{slicing criterion} is defined as a set of neurons with special meanings (such as the neurons in the last layer of an image-classification model whose outputs represent the probabilities of categories), while \textit{a neural network slice} is defined as a subset of neurons in the neural network that exhibit larger effects to the slicing criterion.
\sys focuses on dynamic slicing in which a slice is corresponding to a set of input samples, rather than static slicing, which is input-independent.

\sys consists of three phases: a profiling phase, a forward analysis phase, and a backward analysis phase.
The profiling phase aims to model the average behavior of each neuron. The behavior of a neuron can be characterized by its activation values, which changes by feeding different data samples into the model. We feed all training data into the model and compute the average activation value of each neuron . These average values are used as the baseline to understand the reaction of each neuron to specific data samples.

The forward analysis feeds the interested data samples (the samples we want to compute slice with) into the model and records the activation value of each neuron. The difference between the recorded value and the average activation value computed in the profiling phase represents the neuron reaction to the data samples. The magnitude of the value difference indicates the sensitivity of the neuron with regard to the data samples.

However, the neurons with higher sensitivity are not necessarily more important for the slicing criterion, since the effect of the neuron may be eliminated or redirected to other outputs by its subsequent neurons.
Thus, we further perform a backward analysis that backtracks the data flow from the output neurons to understand the contribution of each neuron.
Specifically, the slice is initialized with the output neurons specified in the slicing criterion. We then iteratively analyze each neuron in the slice by calculating the contributions of its preceding neurons. The preceding neurons with higher contributions are added into the slice for further backward analysis.



We implement \sys in TensorFlow through instrumentation and support the common operators in convolutional neural networks (CNNs). Our implementation is able to deal with large state-of-the-art CNN models, such as ResNet \cite{he2016deep}. The time spent by \sys to compute a slice for a data sample is around 40 seconds on ResNet10 and 550 seconds on ResNet18. Computing slices for batch input is much faster (about 3s and 40s per sample on ResNet10 and ResNet18).

To demonstrate the usefulness and effectiveness of \sys, we further build three applications for adversarial defense, model pruning and model protection, respectively.
First, we show that \sys can be used to effectively detect adversarial samples. Specifically, we show that the slice computed for a data sample reflects how the prediction decision is made by the model, and the slices computed from adversarial samples significantly differ from the slices computed from normal samples. On average, the adversarial input detector implemented based on \sys achieves a high precision of 0.83 and a perfect recall of 1.0.
Second, we show that \sys can be used to customize DNN models for a certain label space. Given a subset of model outputs, \sys computes a slice for the outputs and generates a smaller model that is composed of the neurons and synapses in the slice. We show that the sliced model significantly outperforms other model-pruning methods. Notably, the sliced model can achieve high accuracy (above 80\%) even without fine-tuning.
Finally, \sys can also be used to improve model protection. Specifically, we can selectively protect the important slices rather than the whole model, in order to reduce the protection overhead. We show that by hiding 50\% parameters selected by \sys, the exposed part can be nearly immune to model extraction attacks \cite{orekondy2019knockoff}.


This paper makes the following contributions:

\begin{enumerate}
    \item To the best of our knowledge, this is the first paper to systematically explore and study the idea of dynamic DNN slicing.
    \item We implement a tool, \sys, for dynamic DNN slicing on the popular deep learning framework TensorFlow. Our tool is scalable and efficient.
    \item We develop three interesting applications using DNN-slicing techniques and demonstrate the effectiveness of \sys.
\end{enumerate}

\section{Background and related work}

\subsection{Deep Neural Networks}

Deep neural networks (DNNs) are inspired by the biological neural networks that constitute animal brains. A neural network is based on a collection of connected mathematical operation units called artificial neurons. Each connection (synapse) between neurons can transmit a signal from one neuron to another. The receiving neuron can process the signal(s) and then signal downstream neurons connected to it.
Typically, neurons are organized in layers. Different layers may perform different kinds of transformations on their inputs.
For a certain kind of neuron, how it processes the signal is determined by its weights, which are learned by considering examples. For example, in image recognition, the neural network learns from example images that have been manually labeled as "cat" or "no cat" and uses the learned knowledge to identify cats in other images. Neural networks are good at capturing complex mapping relations between inputs and outputs that are difficult to express with a traditional rule-based program.
Today, DNNs have been used on a variety of tasks, including computer vision \cite{cirecsan2012multi}, natural language processing \cite{gers2001lstm}, recommendation systems \cite{elkahky2015multi}, and various software engineering tasks \cite{gu2016deep_api,li2019humanoid}, where they have produced results comparable to and in some cases superior to human experts.

A simple neural network is shown in Figure~\ref{figure/overview} (a). The neural network contains 9 neurons (2 input neurons, 2 output neurons and 5 intermediate neurons organized in 2 hidden layers) and 16 synapses. The first hidden layer contains 3 neurons that receive signals from the input neurons and send signals to the second hidden layer, which contains 2 neurons that further process the signals and forward to the output neurons. In this example, each neuron (except the input neurons) performs a weighted sum operation, which multiplies each received signal with a learned weight (marked on the synapses) and computes the sum as the neuron's value. Such weighted sum operations are common in today's neural networks, while usually accompanied by other operations such as rectifier, maximum, etc. The example neural network is for illustration purpose only and does not produce meaningful output. Real-world deep neural networks typically have millions of neurons and synapses \cite{he2016deep,szegedy2015going,devlin2018bert}.

\subsection{Program Slicing}

Program slicing is a fundamental technique to support various software engineering tasks in traditional programs, such as debugging, testing, optimization, and protection.
It was originally introduced by Mark Weiser in 1981 \cite{weiser1981slicing} for imperative, procedureless programs. It aims to compute a program slice $S$ that consists of all statements in program $P$ that may affect the value of variable $v$ in a statement $x$. The slice is defined for a slicing criterion $C=(x,v)$ where $x$ is a statement in program $P$ and $v$ is variable in $x$. The slicing criterion represents an analysis demand relevant to an application, \eg, in debugging, the criterion could be the instruction that causes a crash.

At first, only static program slicing was discussed, which analyzes the source code to find the statements that can affect the value of variable $v$ at statement $x$ for any possible input.
Korel and Laski \cite{korel1988dynamic} introduced the idea of dynamic program slicing, which tries to find the statements that actually affect the value of a variable $v$ for a particular execution of the program rather than all statements that may have affected $v$ for any arbitrary execution of the program. 

Program slicing techniques have been seeing a rapid development since its original definition.
Various approaches are proposed to improve the slicing algorithms \cite{horwitz1990interprocedural,zhang2003precise}, introduce other forms of slicing \cite{harman2003amorphous,canfora1998conditioned} and extending slicing ability to more programming languages and platforms \cite{azim2019dynamic,uzuncaova2007kato,binkley2014orbs,clarke1999program}.
Meanwhile, many applications of program slicing techniques are proposed. Today, program slicing is widely used in various software engineering tasks including debugging \cite{agrawal1993debugging}, testing \cite{arlt2014reducing}, software verification \cite{clarke1999program}, software maintenance \cite{gallagher1991using}, and privacy analysis \cite{li2017privacystreams}.
There are many comprehensive surveys \cite{tip1994survey,binkley2004survey,silva2012vocabulary} that summarize the advances in this area.


In this paper, we try to implement program slicing for deep neural networks, a completely different type of program that consists of mathematical operations with learned weights, rather than developer-written statements or variables.

\subsection{Program Analysis for Neural Networks}


Prior to ours, researchers had already attempted to analyze neural networks by applying or borrowing ideas from traditional program analysis techniques.

One of the most widely discussed applications of neural network analysis is to test the robustness of neural networks against adversarial attacks \cite{szegedy2013intriguing,zhang2020mltesting,li2020learning}, which add small perturbation to the input to fool the DNN models. DeepXplore \cite{pei2017deepxplore} proposed to use neuron coverage (the number of activated neurons) to measure the parts of a deep learning system exercised by a set of test inputs, and higher coverage usually means higher robustness. Since then, several new coverage metrics were introduced and various approaches were proposed to generate test inputs that maximize the coverage \cite{tian2018deeptest,xie2019deephunter,du2019deepstellar}. Training the model with the generated test inputs can improve its robustness and accuracy.

In addition to testing, many studies have attempted to detect adversarial inputs based on the internal behavior of neural networks. For example, Gopinath~\etal \cite{gopinath2019property} and Ma~\etal \cite{ma2019nic} attempted to extract properties or invariants from the neuron activation state and use them to detect adversarial inputs. Wang~\etal \cite{wang2019model_mutation} borrowed the idea of mutation testing and found that adversarial samples are usually more sensitive to model mutations. Qiu~\etal \cite{qiu2019adversarial} and Wang~\etal \cite{wang2018interpret} extracted a path from the neural network that is the most critical for a sample, which can be used to distinguish normal and adversarial samples.
The slice computed in our approach can also be viewed as the decision logic of the neural network and used to identify adversarial samples (discussed in Section~\ref{section:defense}).

As neural networks are inherently vulnerable and imprecise, researchers had also tried to provide a formal guarantee of security and correctness with the help of program analysis techniques, such as constraint solving \cite{katz2017reluplex}, interval analysis \cite{wang2018formal,wang2018neurify}, symbolic execution \cite{gopinath2018symbolic}, and abstract interpretation \cite{gehr2018ai2}. While promising, these techniques usually suffer from poor scalability - most of them cannot be applied to today's large neural networks.




There are also some existing work incorporating the idea of ``slicing'' to neural networks.
Shen~\etal \cite{shao2016slicing} proposed slicing CNN feature maps to understand the appearance and dynamic features in the videos. 
Cai~\etal \cite{cai2019model} proposed to slice a DNN into different groups that can be assembled elastically to support dynamic workload.
However, these approaches are not related to program slicing that aims to understand the internal logic of a program. Instead, they focused on training or assembling different parts of a DNN. 

Qiu~\etal's work~\cite{qiu2019adversarial} is the closest to ours. Given an image classification model, they compute an effective path for each class, which contains the neurons and synapses that positively affect the prediction result. However, with regard to the slicing criterion, their effective paths may be incomplete (\ie, missing important neurons and synapses such as the ones with negative contributions) and imprecise (\ie, including less important neurons and synapses such as the ones yielding a large value for any input). Such shortcomings make their method less useful on applications other than adversarial defense (details in Section~\ref{section:applications}).

\section{Motivation and goal}

\subsection{Motivation}

Similar to traditional programs, we argue that slicing a DNN is also meaningful and useful for many important software engineering tasks, as illustrated below.

First, a DNN is a black box whose decisions are hard to interpret \cite{zhang2018visual}. As a result, it is usually hard or even impossible for developers to understand when and why a DNN makes mistakes. As in traditional programs, the input would take a different control flow or data flow if it leads to failures. It would be potentially beneficial if there is a technique to automatically analyze the decision logic in DNNs.

Second, the size of state-of-the-art DNNs and their required computing power have been growing rapidly in recent years, thus it is highly desirable to reduce the size of DNNs to improve efficiency without sacrificing too much accuracy. 
Model pruning (removing some neurons and synapses) is one of the most widely-used techniques \cite{han2015pruning}. However, how to prune the model (\ie which neurons and synapses to remove) is a key question, as we do not want to remove the critical structures that may lead to severe performance degradation. Deciding which neurons and synapses to prune is quite similar to computing a program slice.

Third, model protection, \ie preventing the model from getting stolen, is on increasing demand as models are traded and shared across different organizations. Various techniques such as homomorphic encryption~\cite{chou2018faster} and hardware enclave \cite{tramer2018slalom} can be used to protect models, but protection often brings performance degradation. A practical solution is to protect a part of the model instead of the whole model \cite{tramer2018slalom}. Thus, partitioning the neural network to important and unimportant slices may be beneficial as we can assign limited protection resources to more important slices.

The similarity between these tasks is the demand to find a subset of neurons and synapses that are more important in the decision-making process, which is the goal of this paper.

\subsection{Problem Formulation}

This section defines the concepts and symbols that will be used in this paper and formulates the goal of DNN slicing.


We first formulate the definition of neuron and synapse, two key concepts used throughout this paper. A neuron $n$ in a neural network is a mathematical operator that takes one or more numerical inputs and yields one numerical output. $n$ is said to be activated if its mathematical operation is executed, and the operation result $y$ is called the activation value. A neuron $n$ has one or more synapses $s_1, s_2, ..., s_k$, weighted with $w_1, w_2, ..., w_k$, respectively. Each synapse $s_i$ scales the activation value of another preceding neuron $x_i$ with $w_i$ and passes the scaled value to the neuron $n$ as input. Similarly, the activation value of neuron $n$ is also passed to other succeeding neurons by other synapses. The very last neurons that do not have succeeding neurons are the output neurons, whose activation values are the output of the neural network model.

Any modern DNN architecture can be viewed as a combination of such neurons and synapses. 
For example, a fully connected layer that maps 20 inputs to 10 outputs can be seen as a combination of 10 neurons, each of which computes the sum of values from 20 weighted synapses. A $16 \times 3 \times 3 \times 32$ filter in a convolutional layer can be viewed as 32 neurons, each of which computes the sum over 144 weighed synapses. A Rectified Linear Unit (ReLU) can be viewed as a neuron with only one synapse.
Note that a neuron may be activated several times with different input values during the inference pass of a sample, such as the neurons in convolutional layers.

\begin{table}
\caption{Definition of symbols commonly used in this paper.}
\label{table:symbol}
\small
\begin{tabular}{cc}
\toprule
Symbol & Meaning \\
\midrule
$\mathcal{M}=(\mathcal{N}, \mathcal{S})$ & \tabincell{c}{Model $\mathcal{M}$ with neuron set $\mathcal{N}$ and synapse set $\mathcal{S}$} \\
\hline
$n, y$ & Neuron $n$ and its activation value $y$ \\
\hline
$s, x, w$ & Synapse $s$, its input value $x$ and weight $w$ \\
\hline
$\mathcal{I}, \xi$ & Input dataset $I$ and an input sample $\xi \in \mathcal{I}$\\
\hline
$\mathcal{O}, o$ & Output neuron set $\mathcal{O}$ and an output neuron $o \in \mathcal{O}$ \\
\hline
$\mathcal{C}, \mathcal{M}^\mathcal{C}$ & Slicing criterion $\mathcal{C}$ and its corresponding slice\\
\hline
$CONTRIB$ & \tabincell{c}{Cumulative contribution of a neuron or a synapse \\ \ie the contribution to the slicing criterion} \\
\hline
$contrib$ & \tabincell{c}{Local contribution of a neuron or a synapse \\ \ie the contribution generated in an operation} \\
\hline
$\theta$ & Hyperparameter to control the slice quality \\
\bottomrule
\end{tabular}

\end{table}

Based on the concept of neurons and synapses, we further define the symbols that will be commonly used later, as shown in Table~\ref{table:symbol}. The formal definition of neural network slicing is given as follows:
\begin{defn}
\textbf{(Neural network slicing)}
Let $\mathcal{M}=(\mathcal{N},  \mathcal{S})$ represents a neural network and $\mathcal{C}=(\mathcal{I}, \mathcal{O})$ is a slicing criterion.
$\mathcal{I}=\xi_1, \xi_2, \dots, \xi_n$ is a set of model input samples of interest and $\mathcal{O} = o_1, o_2, \dots, o_k$ is a set of $\mathcal{M}$'s output neurons of interest. The goal of slicing is to compute subsets $\mathcal{N}_{\mathcal{C}} \subset \mathcal{N}$ and $\mathcal{S}_{\mathcal{C}} \subset \mathcal{S}$ with respect to $\mathcal{C}$, denoted as $\mathcal{M}_{\mathcal{C}} =  (\mathcal{N}_{\mathcal{C}}, \mathcal{S}_{\mathcal{C}})$, that significantly (above a predetermined threshold) contributes to the value of any output $o \in \mathcal{O}$ for any input sample $\xi \in \mathcal{I}$. 
\end{defn}


\subsection{Challenges}
There are three main challenges to slice a neural network.

\begin{enumerate}
    \item \textbf{Understanding the behavior of each neuron.} Unlike an instruction or a function in traditional programs, a neuron is typically a simple mathematical operation that does not have any high-level semantic meaning. The weights of all neurons in a model are learned as a whole to fit the training data, while each neuron is just a small building block whose functionality is vague. However, to compute a slice, we must first be able to differentiate the neurons based on their behavior. 
    \item \textbf{Quantifying the contribution of each neuron.} In traditional program slicing, each instruction's contribution to the slicing criterion is binary: an instruction either affects or is irrelevant to the values of the criterion. In neural network slicing, almost all neurons are connected to the output neurons in the slicing criterion and contribute to the outputs more or less. It is difficult to quantify the contribution of each neuron  to extract the most important neurons.
    \item \textbf{Dealing with large models.} Today's state-of-the-art neural networks typically contain millions of neurons that are densely connected. Analyzing a network on such a scale poses a higher demand for efficiency. How to design algorithms that can leverage existing computing resources to speed up the analysis is also a challenging problem.
\end{enumerate}

\section{Our approach: \sys}


We introduce \sys to address the above challenges.
Section~\ref{section:overview} presents an overview of our approach.
Section~\ref{section:profiling} describes how we understand neuron behaviors through differential analysis.
Section~\ref{section:backward} introduces our backward data flow analysis technique that quantifies the contribution of each neuron to the slicing criterion.
Finally, Section~\ref{section:acceleration} briefs how the computation power of GPUs and multi-core CPUs are utilized to improve the efficiency of our method.

\subsection{Approach Overview}
\label{section:overview}

\begin{figure*}[t]
\centering
\includegraphics[width=14.6cm]{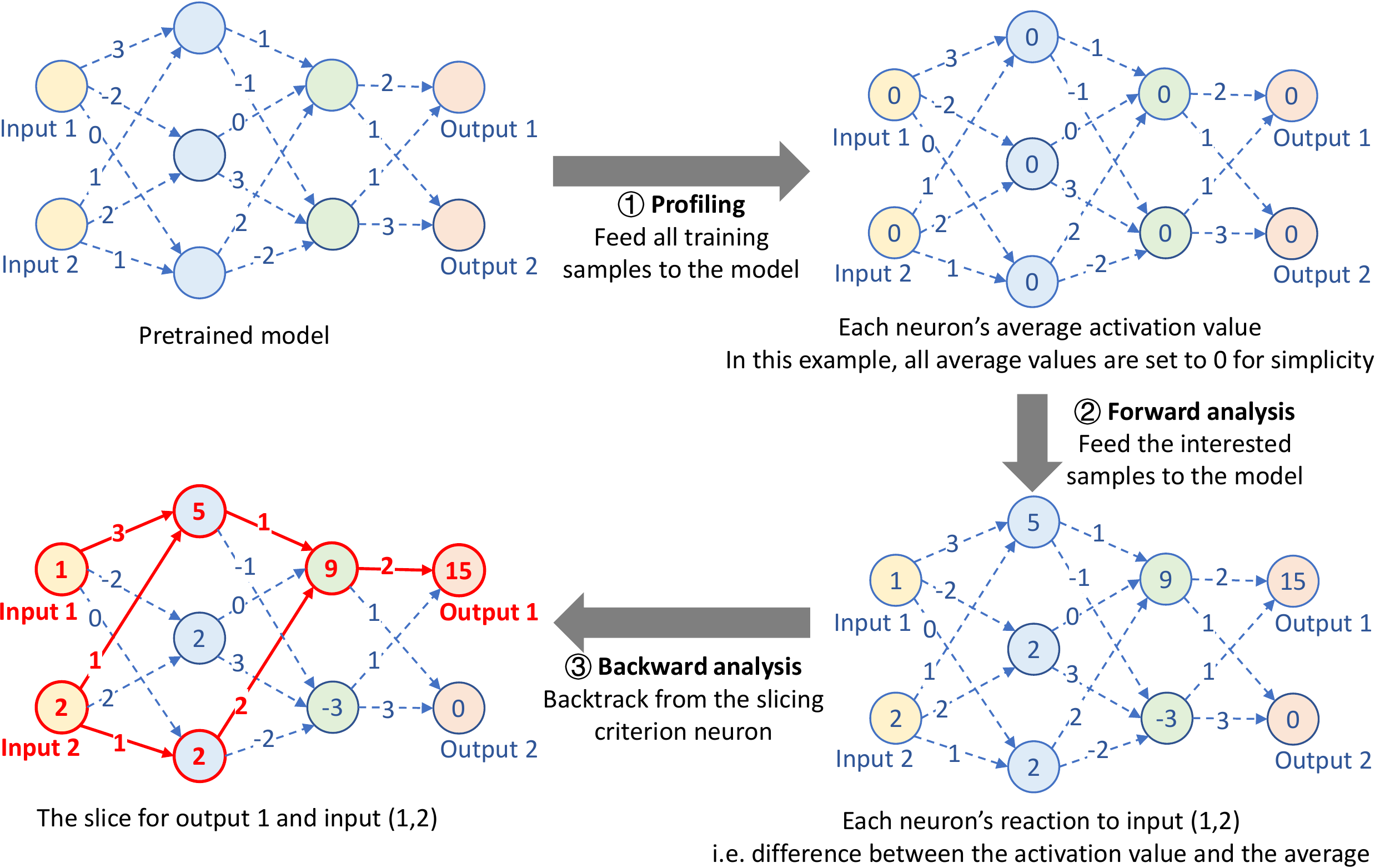}
\caption{The overview of our approach.}
\label{figure/overview}
\end{figure*}

The overview of our approach is illustrated in Figure~\ref{figure/overview}. The program under analysis in our system is a pretrained neural network model, whose weights are already learned to fit a training dataset.
In Figure~\ref{figure/overview}(a), the weight values are labeled on the corresponding synapses in the network.
Our approach mainly consists of three phases, including a profiling phase, a forward analysis phase, and a backward analysis phase.

In the profiling phase, all samples in the training dataset are fed into the model, each sample produces an activation value at each neuron. We log the activation values of each neuron for all input samples and compute the mean activation value, which is the output of the profiling phase (as labeled on each neuron in Figure~\ref{figure/overview}(b)). The mean activation values can be viewed as the behavioral standard of a neuron.
Then, in the forward analysis phase, each interested sample in the slice criterion is fed into the model. We record the activation value of each neuron  and compute its difference with the mean activation value obtained through profiling (as labeled on each neuron in Figure~\ref{figure/overview}(c)). Such relative activation values represent the neuron reaction to the input sample.
Finally, in the backward analysis phase, we start from the output neurons defined in the slicing criterion and iteratively compute the contributions of preceding synapses and neurons. The synapses and neurons with larger contributions are the slices computed for the slicing criterion.
Each step is detailed and formulated in the following sections.


\subsection{Profiling and Forward Analysis}
\label{section:profiling}

The behavior of a neuron during an inference pass is represented as an activation value (or a list of activation values if the neuron was activated several times). The activation values are arbitrary numbers produced by simple mathematical operations. We first need to make sense of the activation values. Specifically, does the neuron react positively or negatively, and how much?

Our method is inspired by the work on differential power analysis \cite{kocher1999differential}, which decodes the power consumption measurements of a circuit by testing the circuit with different inputs. The power consumption difference can be used to infer the input and program logic.
In our case, the activation value of a neuron is like the power consumption measurement that barely makes sense by itself, but the difference between the activation values for different input samples can reveal how the neurons react to each sample.

Specifically, we use the difference between \emph{the neuron behavior for an input sample} and \emph{its average behavior for all training samples} to understand the neuron reaction to the input.
Suppose $\xi$ is an input sample and $n$ is a neuron of model $\mathcal{M}$. By feeding $\xi$ into $\mathcal{M}$, we would observe an activation value $y^n(\xi)$ at neuron $n$. $y^n(\xi) = {mean}_{i=1}^m {y^n_i(\xi)}$ if $n$ is activated multiple times, where $y^n_i(\xi)$ is the $i$-th activation value and $m$ is the total number of activations of $n$ (\eg $m=1$ if $n$ is a neuron in a fully connected layer, and $m$ equals to the number of convolution operations performed by the filter if $n$ is in a convolutional layer).
The average neuron activation value over the whole training dataset $\mathcal{D}$ is calculated by:
\begin{equation}
\label{equation:average_activation}
\overline{y^n(\mathcal{D})} = \frac{\sum_{\xi \in \mathcal{D}} y^n(\xi)}{|\mathcal{D}|}
\end{equation}

Such average activation values can be viewed as the behavioral standard of the neurons, which can be used as the baseline to measure a neuron's reaction to a specific data sample.
Since $\overline{y^n}$ is not dependant on any specific input or output, it only needs to be computed once and can be used for different slicing goals.

In the forward analysis phase, we quantify the reaction of the neuron $n$ for a specific data sample $\xi$ as its relative activation value:
\begin{equation}
\label{equation:neuron_reaction}
\Delta y^n(\xi) = y^n(\xi) - \overline{y^n(\mathcal{D})}
\end{equation}

A positive $\Delta y^n(\xi)$ means that neuron $n$ reacts more positively to the sample $\xi$ than most other samples, and vice versa. The magnitude of $|\Delta y^n(\xi)|$ represents the sensitivity of $n$ with regard to $\xi$. As an example, the output neuron of an image classification model that is trained to detect cats would be more sensitive and positively react to an image of a cat, as compared with an image of a truck.



\subsection{Backward Analysis and Slice Extraction}
\label{section:backward}

The backward analysis aims to compute the contribution of each neuron and each synapse to the interested outputs in the slicing criterion.
Note that the neuron's reaction to an input sample computed through the profiling and forward analysis is not equivalent to its contribution.
For example, in an image classifier, a neuron that reacts sensitively to cat images may not have any contribution if our interested output is the ``truck'' label.
To compute the contribution, we introduce a backward data flow analysis method.


In traditional programs, extracting the instructions and variables that contribute to a certain output is easy based on the data flow graph (DFG), which defines the data dependencies between the instructions and variables. A neural network can also be viewed as a data flow graph, but the graph is densely connected. For modern DNNs that are organized layer by layer, almost every neuron in one layer is connected to all neurons in the previous layer (as shown in Figure~\ref{figure/overview}). Thus, we need to further analyze the data flow graph to measure the contribution of each neuron.

The contribution of a neuron or synapse is quantified as an integer in \sys, denoted as $CONTRIB$. $n$ is a critical neuron if $CONTRIB_n \neq 0$, and a critical neuron may contribute positively ($CONTRIB_n > 0$) or negatively ($CONTRIB_n < 0$) to the slicing criterion. The same for the synapses.

Our method to compute $CONTRIB$ is to recursively compute the contributions of preceding neurons and synapses  from back to front. 
Given a neural network and a list of target neurons, we first consider the neurons that are directly connected to the interested neurons, whose contribution can be extracted with their activation values (in detail later). Then we remove the target neurons from the network and set the neurons with non-zero contribution as the target neurons. We repeat the process until the target neurons do not have any neighboring neurons. 
The algorithm is described in Algorithm~\ref{algorithm:backward}. Note that in practice neurons are organized as partially ordered layers, thus each iteration of Algorithm~\ref{algorithm:backward} deals with a single layer.

\begin{algorithm}[t]
\caption{\textbf{ComputeContrib}: Computing the contributions of neurons and synapses to a list of target neurons for an input sample}
\label{algorithm:backward}
\begin{algorithmic}[1]
\REQUIRE A neural network model $\mathcal{M}=(\mathcal{N},\mathcal{S})$, an input sample $\xi$ and a list of target neurons $\mathcal{O}$. A global table $CONTRIB$ that stores the cumulative contribution of each neuron and synapse during the inference pass of $\xi$, initialized to 0.
\STATE Terminate if $\mathcal{O}$ is empty
\STATE Initialize $\mathcal{O}' = \emptyset$
\FOR{each neuron $o \in \mathcal{O}$}
    \STATE Find $o$'s preceding neurons and synapses $(N', S')$
    \STATE Compute local contributions of $N'$ and $S'$ as $contrib$
    \STATE Update $CONTRIB$ with $contrib$
\ENDFOR
\FOR{each neuron $n \in \mathcal{N}$}
    \STATE Add $n$ to $\mathcal{O}'$ if $n$ is a predecessor of $\mathcal{O}$ and $CONTRIB_n \neq 0$
\ENDFOR
\STATE Obtain $\mathcal{N}'$ by removing neurons in $\mathcal{O}$ from $\mathcal{N}$
\STATE Call \textbf{ComputeContrib} by setting $\mathcal{O} = \mathcal{O}'$ and $\mathcal{N} = \mathcal{N}'$
\RETURN The global cumulative contribution table $CONTRIB$
\end{algorithmic}
\end{algorithm}

Algorithm~\ref{algorithm:backward} simplifies the problem of \emph{computing cumulative contributions of all neurons and synapses in the whole network} to \emph{computing local contributions of preceding neurons and synapses in an operation} (line 5). Local contributions mean the contributions generated solely by the operation. We use the weighted sum operator (a common operator in neural networks) to illustrate how we compute the contributions of preceding neurons and synapses.

In the weighted sum operator, the central neuron $n$ has $k$ synapses ($s_i, s_2, ..., s_k$) that connect $k$ preceding neurons ($n_1, n_2, ..., n_k$) to $n$. The activation value of $n$ is computed as $y=\sum_{i=1}^{k} w_i x_i$, where $w_i$ is the weight of synapse $s_i$ and $x_i$ is the activation value of $n_i$. Suppose the cumulative contribution of $n$ is $CONTRIB_n$, the local contribution $contrib_i$ of $n_i$ and $s_i$ is computed as:
\begin{equation}
\label{equation:local_contrib}
contrib_i = CONTRIB_n \times \Delta y^n \times w_i \Delta x_i
\end{equation}
in which $\Delta y^n$ is the relative activation value of the central neuron given by Equation~\ref{equation:neuron_reaction} and $\Delta x_i$ is the relative activation value of the neuron $n_i$ (\ie $\Delta y^{n_i}$). The product of $\Delta y^n$ and $w_i \Delta x_i$ represents the impact that $n_i$ and $s_i$ may have on the global contribution $CONTRIB_n$. For example, if $\Delta y^n$ is negative and $w_i \Delta x_i$ is positive, it means that $n_i$ enlarges the negativity of $n$, yielding an contribution that is opposite to $CONTRIB_n$.

\begin{table*}[]
  \caption{Neuron operations considered in \sys.}
  \label{table:ops}
\renewcommand{\arraystretch}{1.5}
  \begin{tabular}{cp{3.5cm}lll}
  \toprule
  Operation & Usage & Math form & Local contribution of $i$-th input \\\hline 
  \textbf{Weighted sum} & {Convolutional layers and fully connected layers, etc.} & $y=\sum\limits_{i=1}\limits^{k} w_i x_i$ & 
  \tabincell{c}{
  $CONTRIB_n \times \Delta y \times w_i \Delta x_i$
  }\\ \hline
  \textbf{Average} & {Average-pooling layers.} & $y=\frac{1}{k} \sum\limits_{i=1}\limits^{k} x_i$ & 
  \tabincell{c}{
  $CONTRIB_n \times \Delta y \times \Delta x_i$
  }\\ \hline
  \textbf{Maximum} & \parbox[c]{3cm}{Max-pooling layers.} & $y=\max_{i=1}^{k} x_i$ & 
  $CONTRIB_n \times \Delta y \times \Delta x_i\ if\ x_i = y \ else\ 0$
  \\ \hline
  \textbf{Rectify} & \parbox[c]{3cm}{ReLU activation.} & $y=x \; \text{if} \; x > 0 \; \text{else} \; 0$ & 
  $CONTRIB_n \times \Delta y \times \Delta x\ if\ x > 0\ else\ 0$
  \\ \hline
  \textbf{Scale} & \parbox[c]{3cm}{Batch normalization.} & $y=\frac{x-\mu}{\sigma}$ & 
  $CONTRIB_n \times \Delta y \times \Delta x$
  \\
  \bottomrule
  \end{tabular}

\end{table*}

The weighted sum operators take the vast majority in today's DNNs, but there are also other types of operators. In this paper, we focus on convolutional neural networks (CNNs). Table~\ref{table:ops} shows five common operators that are enough to handle most existing CNN models. To support other architectures one only needs to define the method to compute local contributions for new operators, as shown in Table~\ref{table:ops}.

The cumulative contribution $CONTRIB$ of neuron $n_i$ and synapse $s_i$ in the operation is updated by their local contribution:
\begin{equation}
\label{equation:cumulative_contribution}
\begin{split}
CONTRIB_{n_i} += sign(contrib_i) \\
CONTRIB_{s_i} += sign(contrib_i)
\end{split}
\end{equation}
We only keep the sign of the local contribution, as different operations may have different scales of local contributions.

However, updating the cumulative contribution for all neurons and synapses is time-consuming (a neuron with non-zero cumulative contribution introduces a new branch during backtracking) and may accumulate contributions from unimportant neurons and synapses.
Thus, we limit the number of local contributions used to update $CONTRIB$. The importance of a local contribution is represented by its magnitude, and those with smaller magnitude can be excluded when updating $CONTRIB$.
Specifically, we first sort the local contributions in ascending order of their magnitudes. The preceding neurons are sorted as $n_1, n_2, ..., n_k$.
Then we try to find a maximum index $j$ so that $n_1, ..., n_j$ can be excluded while the influence on the activation value of $n$ is below the threshold $\theta$. For example, in a weighted sum operation, the influence of excluding $n_1, ..., n_j$ is $|{\sum_{i=j}^k w_i \Delta x_i}/{y}|$.
$\theta$ controls the amount of excluded local contributions with minimal influence on the functionality of an operation, and thus the generated slice can be directly used to make predictions without retraining (evaluated in Section~\ref{section:prune}). The value of the threshold $\theta$ can be set by different applications to control the size of the resulting slice. 

So far the cumulative contribution $CONTRIB$ captures the contribution of neurons/synapses during the inference of a single input sample. For a slicing criterion $\mathcal{C} = (\mathcal{I}, \mathcal{O})$ that may contain multiple interested samples, the final cumulative contribution is the sum of the contribution for each sample $\xi \in \mathcal{I}$. A slice for $C$ is $\mathcal{M}^C = (\mathcal{N}^C, \mathcal{S}^C)$ where $\mathcal{N}^C$ and $\mathcal{S}^C$ are the neurons and synapses with non-zero contributions. One can also control the size of slice based on the contributions (as in \S\ref{section:prune}).

\subsection{GPU and Multi-thread Acceleration}
\label{section:acceleration}

\sys takes a forward analysis pass and a backward analysis pass for each data sample $\xi \in \mathcal{I}$ when computing the slices. It might be very time-consuming if $|\mathcal{I}|$ is large. Since the process of computing slice for a data sample is independent of each other, we can take advantage of the parallel characteristic of GPU and multi-threading to accelerate the overall slicing process. 

Specifically, for a large set of data samples $|\mathcal{I}|$, we first run the profiling and forward analysis phases on GPU using large batches, as these two phases only involve forward computation. Then a large batch is separated into several small batches. The backward analysis of each small batch runs on the CPU as a separate thread. Finally, the batches are merged together to generate the slicing result.

\section{Implementation \& overhead}
\label{section:evaluation}

\begin{table}
\caption{The time spent to process an input sample in each phase. The profiling and forward analysis phases take the same amount of time as they both only require an inference pass.}
\label{table:evaluation}
\begin{tabular}{cccccc}
\toprule
\multirow{2}*{Model} & \multirow{2}*{\#Params} &  \multicolumn{2}{c}{\tabincell{c}{Profiling/Forward} } & \multicolumn{2}{c}{Backward} \\
 & & Single & Batch & Single & Batch \\
\midrule
LeNet & 42784 & 3.0s & 0.3s & 0.5s & 0.3s \\
\hline
ResNet10 & 300K & 8.9s & 0.4s & 30.1s & 3.0s \\
\hline
ResNet18 & 11M & 9.6s & 0.8s & 543.0s & 40.4s \\
\bottomrule
\end{tabular}	

\end{table}

We implemented \sys in Python with TensorFlow. The profiling and forward analysis are implemented based on TensorFlow's instrumentation mechanism.
The multi-thread computing is implemented by the distributed python library Ray (https://ray.io).

We evaluated the time overhead of \sys on a server that has 2 GeForce GTX 1080Ti GPUs, 2 Intel Xeon CPUs with 16 cores, and 64GB memory.
Table~\ref{table:evaluation} reports the slice time and the architecture complexity of three models. The time spent by \sys to compute slice for a data sample is roughly 4s, 39s, and 553s for LeNet, ResNet10, and ResNet18 respectively. When computing slice for a batch of inputs, the speed is much faster, which is about 0.6s, 3.4s, and 45.2s per data sample for the three models respectively. Note that the profiling phase is not included when computing the slicing speed as it only needs to run once for a model. 

\newcommand{\qiu}{\emph{EffectivePath}\xspace} 
\section{Applications}
\label{section:applications}

In this section, we describe three applications to demonstrate the usefulness and the effectiveness of \sys, including adversarial defense, model pruning, and model protection. In each application, we describe why the application is meaningful, how \sys can help, and how \sys performs compared with other methods.

The main method which we compare \sys against is the state-of-the-art work by Qiu~\etal \cite{qiu2019adversarial} (denoted as \qiu below). We also include some other baselines for more comparisons.

\subsection{Adversarial Defense}
\label{section:defense}

Adversarial examples \cite{szegedy2013intriguing} are carefully-crafted inputs that may lead to wrong predictions. They are usually generated by adding small permutation to a benign input, which is barely noticeable by a human. Adversarial attacks may cause severe consequences, especially in safety- and security-critical scenarios.

As a result, adversarial defense became a hot research topic in both AI and SE communities.
Many approaches tried to make the DNNs more robust through training \cite{zhang2019you, madry2017towards} or adding advanced architectures \cite{ross2018improving, mustafa2019adversarial}, but it is still hard to obtain a 100\% robust model.
Instead, some researchers opted to take another direction: adversarial input detection \cite{ma2019nic,qiu2019adversarial,wang2019model_mutation}, with which, the deep learning system can raise warnings or stop serving once suspicious inputs are detected. Thus, severe attacks can be avoided.
In this section, we discuss how \sys can be used to detect adversarial inputs.


\subsubsection{Method.}

\begin{figure}
    \centering
    \includegraphics[width=0.9\linewidth]{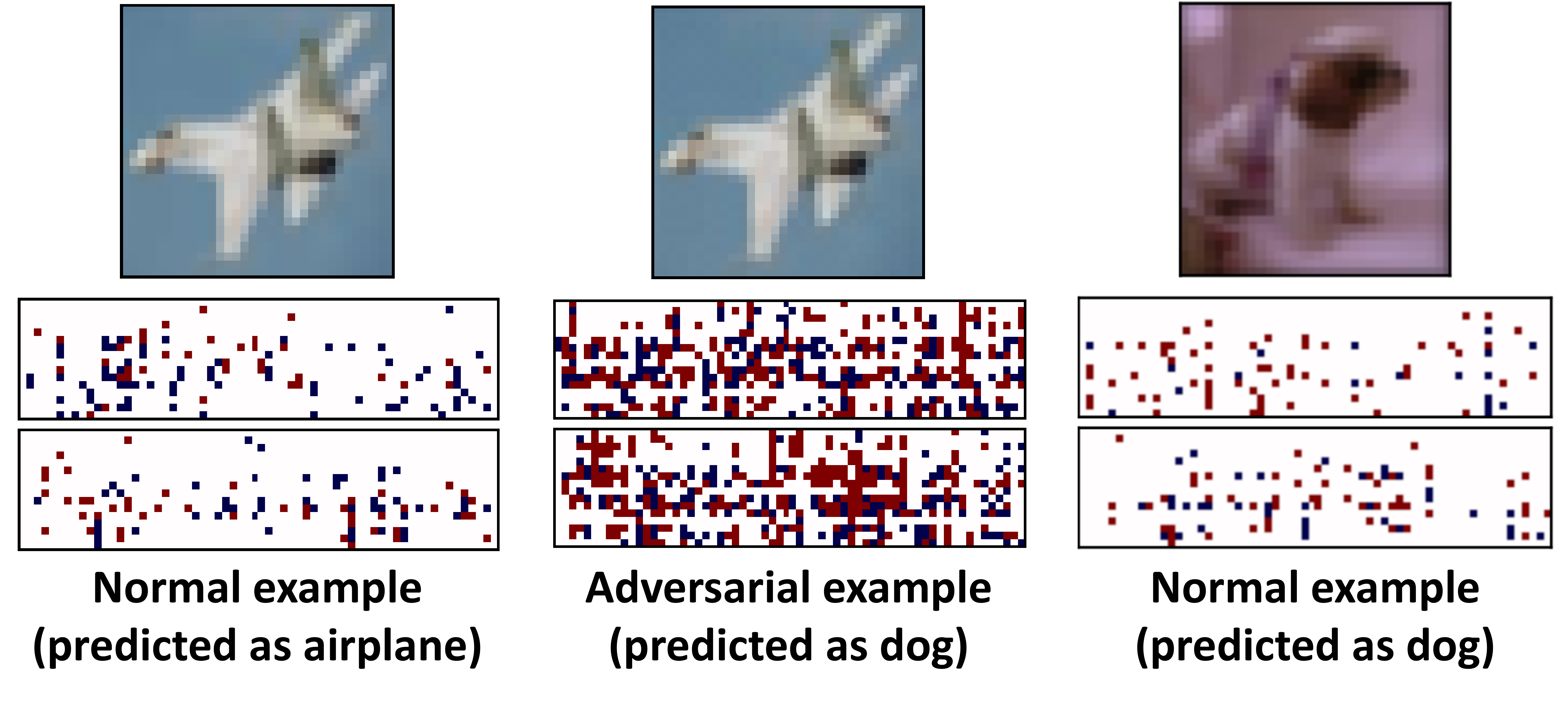}
    \caption{Normal and adversarial examples (top) and their visualized slices (bottom). Each pixel in the visualization represents a neuron from a random convolutional layer (separated to two rows). The neurons with non-zero contributions are colored (blue for neurons with positive contributions and red for those with negative contributions).}
    \label{figure:defense:visualization}
\end{figure}
Our insight is that the slice computed by \sys can be viewed as an abstraction of the decision process, and the decision processes of normal examples and adversarial examples are intuitively different. As shown in Figure~\ref{figure:defense:visualization}, although the normal image and the adversarial one are indistinguishable for a human, their slices are different.
Thus, by learning from the slices of large-scale normal examples, we can understand the normal decision process of the DNN. Therefore, given a new input, if its slice is distinctly different from the normal slices, it is very likely an adversarial input.

Specifically, suppose $\mathcal{M}$ is the DNN model that may accept adversarial inputs, $\xi$ is an input sample and $\mathcal{M}(\xi)$ is the label of $\xi$ predicted by $\mathcal{M}$. Using \sys, we can compute a slice $\mathcal{M}_\xi$ for each input $\xi$ by setting the slicing criterion as $\mathcal{C} = (\xi, \mathcal{M}(\xi))$. We build a slice classifier $F$ that predicts the label of an input $\xi$ based on the slice computed for the input $\mathcal{M}_\xi$. By training $F$ with a large number of normal samples, it can capture the mapping pattern between the slice shape and the corresponding output category. With the trained slice classifier $F$, an input $\xi$ is identified as adversarial if $F(\mathcal{M}_\xi) \neq \mathcal{M}(\xi)$, \ie the prediction made by the slice classifier is different from the prediction of the original DNN model.

The input of the slice classifier, \ie a slice $\mathcal{M}_\xi$, is represented as a vector ${vec}_\xi$. Each element in ${vec}_\xi$ corresponds to a synapse and its value is the contribution of the synapse (as described in Section~\ref{section:backward}). For the simplicity of the input and output representations, many classification algorithms may be used to build the slice classifier. We chose to use 
the decision tree \cite{breiman2017classification} as it is easy to implement and debug.

Applying \sys to adversarial-input detection has three advantages: (1) \sys does not require modifying or retraining the original model, and thus \sys can support any DNN models. (2) \sys can scale up to support large state-of-the-art DNN models, while existing methods like ones by Ma~\etal \cite{ma2019nic} and Gopinath~\etal \cite{gopinath2019property} can only support small models. (3) \sys requires only the normal samples to build the defense, but existing methods \cite{qiu2019adversarial, fidel2019explainability, wang2018interpret, ma2018characterizing} need to train a detector with both normal and adversarial examples. As the attackers can always use new adversarial examples, \sys is a much more realistic solution than those existing methods.



\subsubsection{Evaluation.}

We compare our detection method with two baselines. For a fair comparison, the baseline methods use the same classifier to identify adversarial inputs as ours, while the inputs of the classifier are different. \emph{FeatureMap} is a naive baseline that uses the feature maps of convolutional layers as the inputs of the classifier. \qiu is a more advanced baseline that uses the effective path generated by Qiu~\etal \cite{qiu2019adversarial} to train the classifier. The experiments were conducted on ResNet10 and the CIFAR-10 dataset (image size 32$\times$32). All the classifiers were trained with 10,000 normal samples, using their respective feature extraction methods.

We tested \sys and the two baselines on 17 attacks, covering gradient-based attacks, score-based attacks, and decision-based attacks. These attacks include FGSM~\cite{goodfellow2014explaining} with per-pixel maximum modification of 2, 4 and 8 (relative to 256 and referred to as FGSM\_2, FGSM\_4 and FGSM\_8, respectively), Deepfool~\cite{moosavi2016deepfool} with constrain norm $L_2$ and $L_{\infty}$ (referred to as DeepfoolL2 and DeepfoolLinf), JSMA~\cite{papernot2016limitations} attack, PGD~\cite{madry2017towards} attack with random start and per-pixel maximum modification of 2, 4 and 8 (referred to as RPGD\_2, RPGD\_4 and RPGD\_8), the $L_2$ version of CW attack~\cite{carlini2017towards} (CWL2), ADef attack~\cite{alaifari2018adef}, an attack that just perturbs a pixel (SinglePixel), a greedy local-search attack~\cite{narodytska2016simple} (LocalSearch), a boundary attack~\cite{brendel2017decision} (Boundary), an attack of spatial transformation~\cite{engstrom2017rotation} (Spatial), an attack that performs binary search between a normal sample and its adversarial instance (Pointwise), and an attack that blurs the input until it is misclassified (GaussianBlur). All the attacks are implemented with Foolbox~\cite{rauber2017foolbox}.

For each attack method, we generate adversarial examples from 500 randomly picked normal examples. The examples that successfully mislead the model are fed to the detector with their normal examples.
We compute the precision, recall and F1 score of each detector on each attack, as shown in Table~\ref{table:detection}.

\begin{table*}[h]
\caption{Adversarial input detection accuracy for different attack methods.}
\label{table:detection}

\begin{tabular}{c|c|ccc|ccc|ccc}
\hline
\multicolumn{2}{c|}{\multirow{2}*{Attack Method}} & \multicolumn{3}{c|}{\qiu} & \multicolumn{3}{c|}{\emph{FeatureMap}} & \multicolumn{3}{c}{\sys} \\
\cline{3-11}
\multicolumn{2}{c|}{} & F1 & precision & recall & F1 & precision &recall & F1 &precision & recall \\
\hline
\multirow{11}*{Gradient-based} & FGSM\_2 & 0.82 & 0.69 & \textbf{1.00} & 0.55 & 0.58 & 0.53 & \textbf{0.90} & \textbf{0.82} & \textbf{1.00}  \\
\cline{2-11}
& FGSM\_4 & 0.71 & 0.56 & \textbf{1.00} & 0.55 & 0.60 & 0.52 & \textbf{0.91} & \textbf{0.84} & \textbf{1.00} \\
\cline{2-11}
& FGSM\_8 & 0.88 & 0.79 & \textbf{1.00} & 0.58 & 0.62 & 0.55 & \textbf{0.92} & \textbf{0.84} & \textbf{1.00} \\
\cline{2-11}
& DeepFoolLinf & 0.78 & 0.64 & \textbf{1.00} & 0.68 & 0.68 & 0.68 & \textbf{0.92} & \textbf{0.85} & \textbf{1.00} \\
\cline{2-11}
& DeepFoolL2 & 0.78 & 0.64 & \textbf{1.00} & 0.66 & 0.68 & 0.66 & \textbf{0.92} & \textbf{0.85} & \textbf{1.00} \\
\cline{2-11}
& JSMA & 0.82 & 0.69 & \textbf{1.00} & 0.66 & 0.67 & 0.66 & \textbf{0.92} & \textbf{0.85} & \textbf{1.00} \\
\cline{2-11}
& RPGD\_2 & 0.78 & 0.64 & \textbf{1.00} & 0.59 & 0.63 & 0.56 & \textbf{0.91} & \textbf{0.84} & \textbf{1.00} \\
\cline{2-11}
& RPGD\_4 & 0.82 & 0.69 & \textbf{1.00} & 0.55 & 0.62 & 0.50 & \textbf{0.92} & \textbf{0.85} & \textbf{1.00} \\
\cline{2-11}
& RPGD\_8 & 0.75 & 0.60 & \textbf{1.00} & 0.55 & 0.62 & 0.50 & \textbf{0.92} & \textbf{0.85} & \textbf{1.00} \\
\cline{2-11}
& CWL2 & 0.78 & 0.64 & \textbf{1.00} & 0.66 & 0.68 & 0.66 & \textbf{0.92} & \textbf{0.85} & \textbf{1.00} \\
\cline{2-11}
& ADef & 0.85 & 0.73 & \textbf{1.00} & 0.66 & 0.67 & 0.66 & \textbf{0.91} & \textbf{0.84} & \textbf{1.00} \\
\hline
\multirow{2}*{Score-based} & SinglePixel & 0.67 & 0.50 & \textbf{1.00} & 0.52 & 0.44 & 0.64 & \textbf{0.79} & \textbf{0.65} & \textbf{1.00} \\
\cline{2-11}
& LocalSearch & 0.83 & 0.71 & \textbf{1.00} & 0.67 & 0.67 & 0.68 & \textbf{0.92} & \textbf{0.84} & \textbf{1.00} \\
\hline
\multirow{4}*{Decision-based} & Boundary & 0.82 & 0.69 & \textbf{1.00} & 0.69 & 0.69 & 0.70 & \textbf{0.92} & \textbf{0.85} & \textbf{1.00} \\
\cline{2-11}
& Spatial & 0.78 & 0.64 & \textbf{1.00} & 0.59 & 0.56 & 0.64 & \textbf{0.87} & \textbf{0.77} & \textbf{1.00} \\
\cline{2-11}
& Pointwise & 0.87 & 0.76 & \textbf{1.00} & 0.68 & 0.69 & 0.69 & \textbf{0.92} & \textbf{0.85} & \textbf{1.00} \\
\cline{2-11}
& GaussianBlur & 0.87 & 0.76 & \textbf{1.00} & 0.68 & 0.68 & 0.70 & \textbf{0.92} & \textbf{0.85} & \textbf{1.00} \\
\hline
\multicolumn{2}{c|}{\textbf{Average}} & 0.80 & 0.67 & \textbf{1.00} & 0.62 & 0.63 & 0.62 & \textbf{0.91} & \textbf{0.83} & \textbf{1.00} \\
\hline
\end{tabular}

\end{table*}


According to the experiment result, \sys is very effective in detecting adversarial inputs with an average recall of 100\% and an average precision of 83\%, which means that \sys is able to correctly identify all the adversarial examples generated with these attack methods (no false negative). Meanwhile, most of the inputs identified by \sys are indeed adversarial inputs, while only a few normal samples are misidentified (false positives).
Although \qiu also achieves a perfect recall, its precision is much lower, meaning that the detector may easily misclassify normal samples as adversarial inputs.
The average recall of 63\% in \emph{FeatureMap} represents the feature maps between normal examples and the adversarial examples are barely discriminative. This phenomenon indicates the demand for \sys to explore the mechanism of neural networks.


\subsection{Network Simplification and Pruning}
\label{section:prune}


The size and complexity of DNN models grow rapidly. Although these huge models achieve high scores on complicated datasets, they are cumbersome and slow in real-world, task-specific applications. How to reduce the model size and speed up the computation is crucial to the DNN applications.
 
One acceleration technique is to prune trivial synapses of a large model to generate a light-weight one. With redundant weights trimmed off, the computation of executing the model may be reduced. Existing network-pruning methods focus on reducing the network architecture of models for all the output classes~\cite{liu2018rethinking}. With DNN slicing, \sys enables more flexible network simplification and pruning by focusing on a targeted subset of output classes. That is, for a subset of the original output classes of a model, \sys can decide the proper model slices for the targeted output classes. Thus, \sys can generate a smaller model for the targeted output classes with higher model accuracy. This advantage of \sys is highly desirable in real-world applications that usually deal with a small set of output classes (e.g., classifying only different dogs rather than 1,000 types of animals).

\subsubsection{Method.} 

\sys can pick out neurons and synapses critical to a slice criterion $\mathcal{C}=(\mathcal{I},\mathcal{O})$.
By setting $\mathcal{O}$ to the set of interested target classes, \sys can compute $CONTRIB_s$ for each synapse $s$, which represents the synapse's importance to the target classes.
We can trim out the less important synapses and get a model that still functions on the target classes. 

Specifically, suppose we want to prune $\mathcal{M}$ for target classes $\mathcal{O}^T$ with prune ratio $r$. Let $\mathcal{I}^T$ be the set of data samples belonging to the interested classes. $CONTRIB^T$ is the cumulative contributions computed by \sys, and $CONTRIB^T_s$ is the contribution of synapse $s$.
For each layer $l$, we sort all synapses in the layer $\mathcal{S}_l$ by the ascending order of their contributions magnitudes. The first $r \times |\mathcal{S}_l|$ synapses are pruned, and a neuron is also pruned if its synapses are all pruned.



\newcommand{\targeted}{\emph{\sys (targeted)}\xspace} 
\newcommand{\nontargeted}{\emph{\sys (all)}\xspace} 
\newcommand{\synapse}{\emph{Weight}\xspace} 
\newcommand{\neuron}{\emph{Channel}\xspace} 
\newcommand{\random}{\emph{Random}\xspace} 

\subsubsection{Evaluation.}
To evaluate the ability of \sys to targeted pruning, each of 210 subsets of CIFAR10's 10 output classes is used as the target classes $\mathcal{O}^T$.
\targeted represents to prune synapses according to the contributions computed for the target classes $\mathcal{O}^T$. 
\nontargeted represents to prune according to the contributions computed for all output classes $\mathcal{O}$.
The comparison between \targeted and \nontargeted demonstrates \sys's ability in target classes. We also compare it with several baselines. \qiu represents pruning synapses based on the feature computed in~\cite{qiu2019adversarial}. 
\synapse is based on the absolute synapse weights, where the synapses with the smallest weights are trimmed~\cite{han2015deep_compression}. Similarly, \neuron represents to prune the least important neurons by the average connected weight value~\cite{he2017channel}. Both \synapse and \neuron are widely used techniques in the field of network pruning. 

\begin{figure}
    \centering
    \includegraphics[width=0.9\linewidth]{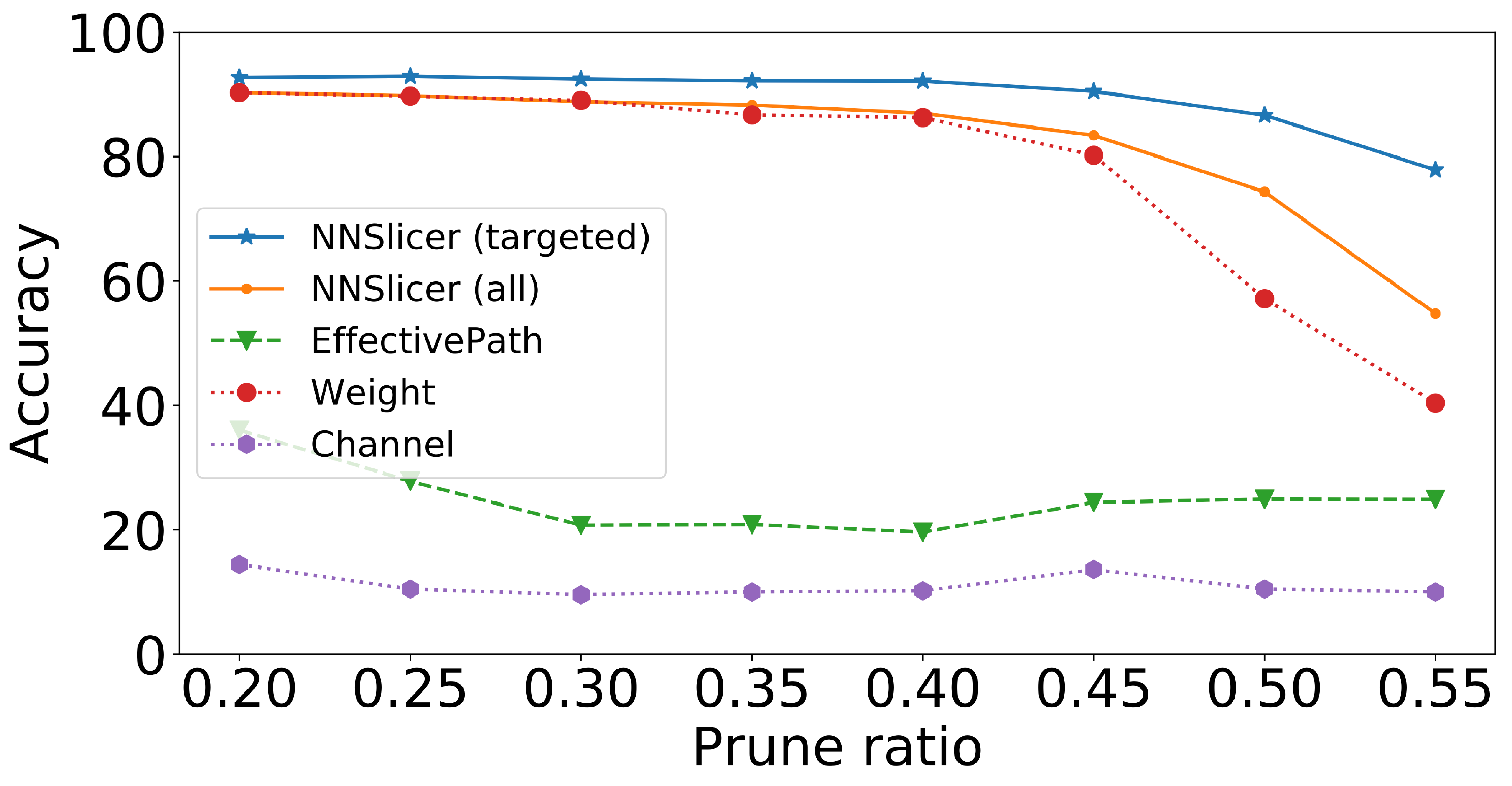}
    \caption{Accuracy of the pruned models without fine-tuning.}
    \label{figure:prune}
\end{figure}

Figure~\ref{figure:prune} shows the average accuracy over possible target classes. The accuracy of \targeted is always high and is around 80\% when 55\% of weights are pruned. 
The accuracy of \qiu and \neuron are both low in the figure. 
The accuracy of \nontargeted and \synapse is high only when the prune ratio is below 45\%.
The comparison between \targeted and \nontargeted demonstrates the ability of \sys to prune for specific classes. 
 The large gap between \targeted and \qiu indicates the advantage of \sys to the feature computed by~\cite{qiu2019adversarial}. 

\begin{figure}
    \centering
    \includegraphics[width=\linewidth]{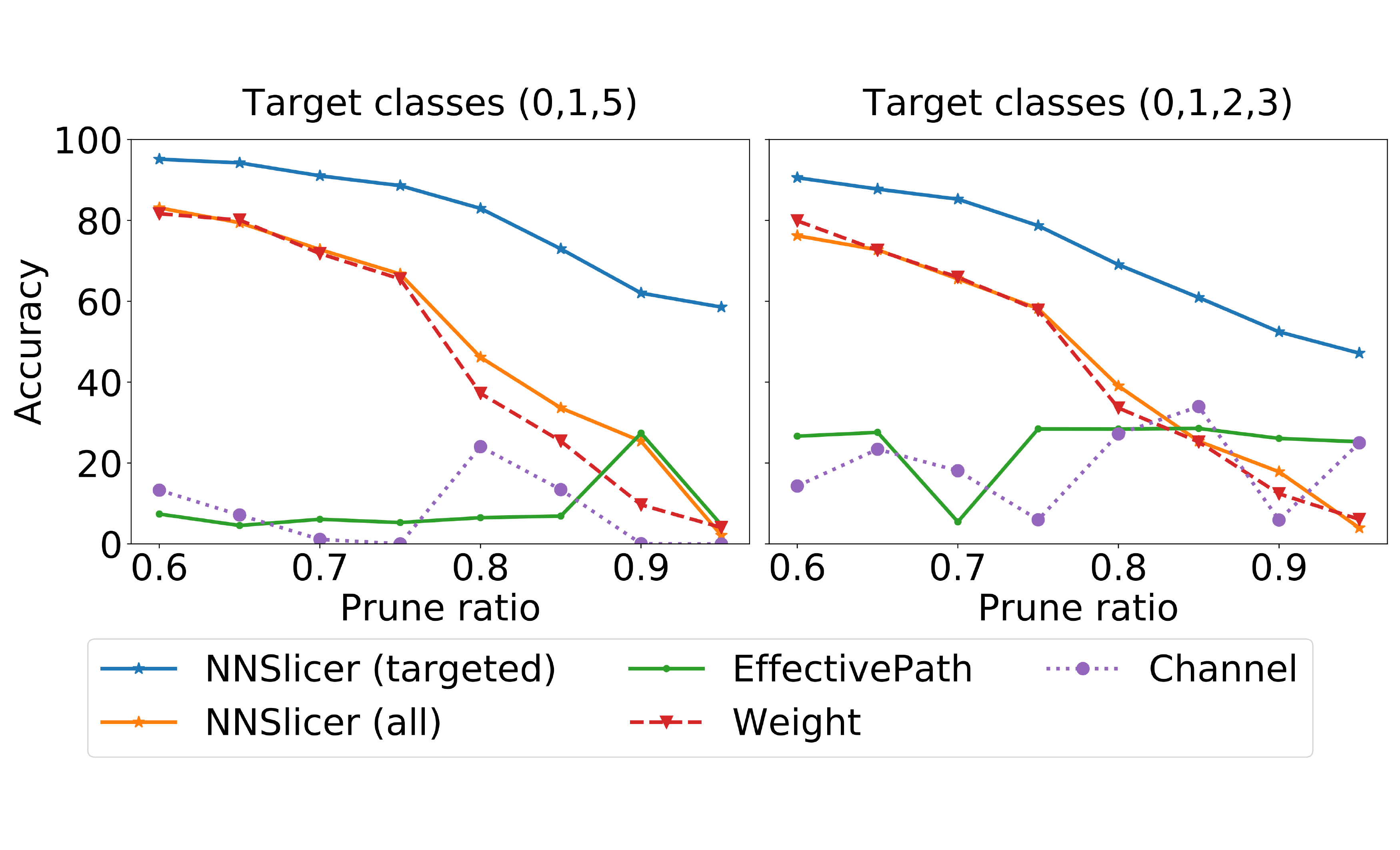}
    \caption{Accuracy of the pruned models after fine-tuning for one epoch.}
    \label{figure:finetune}
\end{figure}

When the prune ratio becomes larger, we further evaluate the performance with fine-tuning. To do it, the pruned models are retrained on 10k samples for 1 epoch. 
Figure~\ref{figure:finetune} shows the performance of the fine-tuned models on two sets of target classes.
The fine-tuned model of \sys is noticeably higher than other methods. It shows that \targeted preserves the model's capability on targets even when a large portion of weights is trimmed. A short fine-tuning (1 epoch in this case) is enough for the model to achieve high accuracy.

One possible reason for \sys's good performance is that it preserves the model's ability to target classes at the cost of other non-target classes. In an extra experiment, the performance of \targeted on non-target classes is remarkably lower than target classes. On the other hand, the difference of \synapse is small.  It means \sys can decompose the model over classes and make a trade-off to conserve the ability on target classes. A similar phenomenon is observed in model protection and will be discussed in Section~\ref{section:protection}.

\subsection{Model Protection}
\label{section:protection}

DNN models are becoming valuable assets due to the high cost of the training process, including collecting a large amount of data, expensive GPU usage, and enormous power consumption. However, an attacker may retain (or steal) the functionality of a model at a comparatively low cost~\cite{correia2018copycat, reith2019efficiently, jagielski2019high, orekondy2019knockoff, hu2019neural}. How to protect models from being stolen is becoming an increasingly important problem, particularly in the emerging edge computing where models are deployed to edge servers or even end devices. 


Existing solutions of model protection usually leverage encryption, using homomorphic encryption~\cite{gilad2016cryptonets, chou2018faster, zhang2019cheetah} or zero knowledge proof~\cite{yampolskiy2011ai}, or running a model inside trusted execution environments~\cite{costan2016intel, tramer2018slalom, costan2016sanctum}. All sensitive computation is conducted in the encrypted mode. However, the cost of these protected computations is high. For example, CryptoNets~\cite{gilad2016cryptonets} takes around 300s to execute a model on the small MNIST dataset. To reduce the cost of model protection, one approach is to secure the important computation only, where \sys may help. 


\subsubsection{Method.}
The existing model protection work is constrained to protecting the model w.r.t. the whole label space~\cite{orekondy2019prediction, juuti2019prada}.
But the importance of outputs may vary. For some outputs, the data is more difficult to collect, or the annotation is particularly more expensive.
Because \sys can slice model for certain classes, it can help to find significant components for the expensive classes and protect them.
We propose to incorporate targeted protection in this scenario. Compared to existing work, our method is more flexible and can customize the protection target.
\sys selects synapses from a model and protects their weights. The way to select synapses is similar to Section~\ref{section:prune} but \sys selects the most crucial synapses for the target classes. The selected synapses are protected from attackers who have to recover the protected synapses through retraining to obtain the whole model.

\begin{figure}
    \centering
    \includegraphics[width=\linewidth]{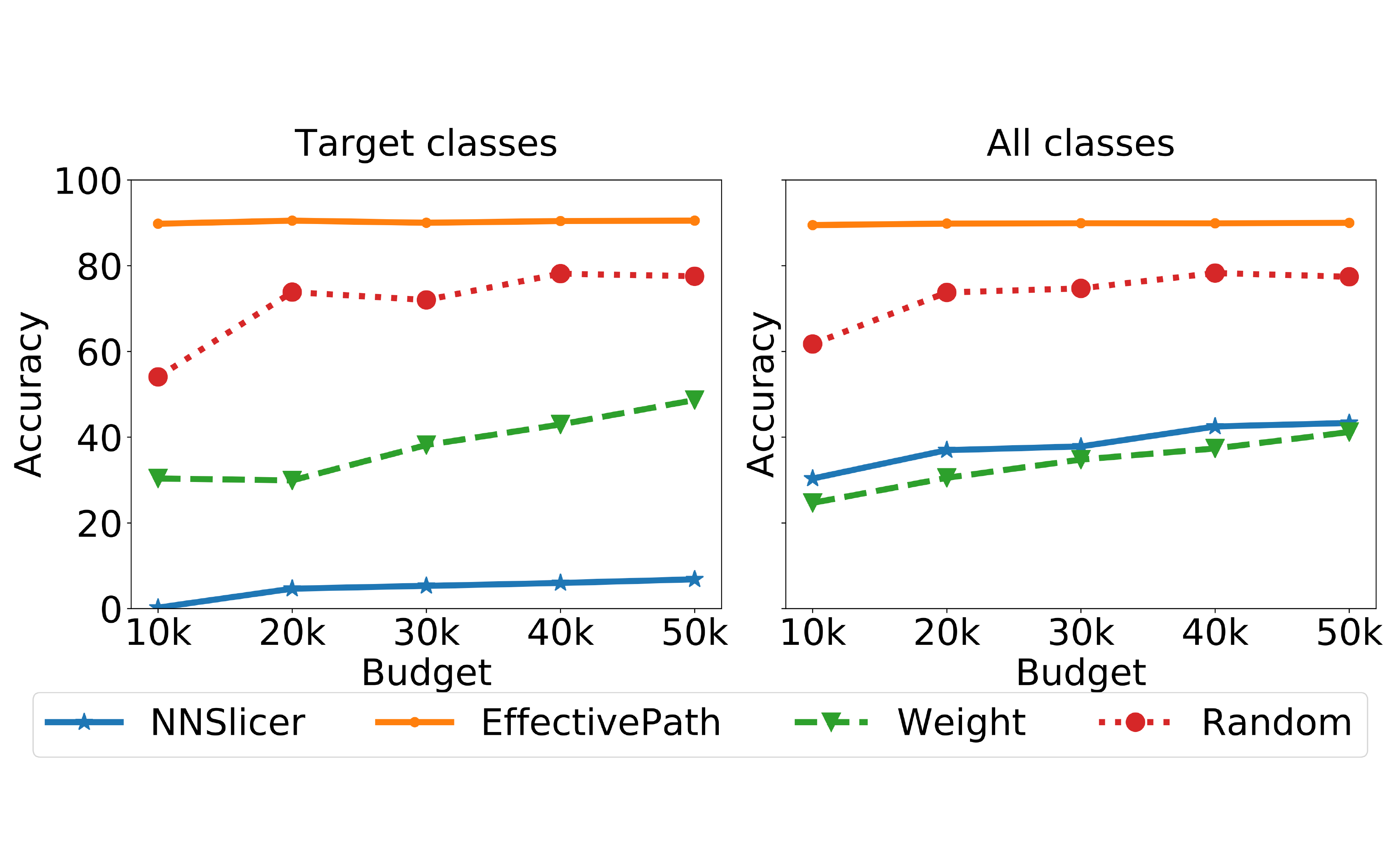}
    \caption{The accuracy achieved by retraining the models for 5 epochs. 50\% of the parameters (selected with different methods) in the model are hidden, and the attacker tries to recover them through retraining. The x-axis is the attacker's budget (\ie number of samples used to retrain). A lower accuracy achieved with a fixed budget means better protection.}
    \label{figure:protect}
\end{figure}








\subsubsection{Evaluation.}
In the experiment, we assume a strong attacker who has a training dataset. The attacker's dataset size is called the budget \cite{orekondy2019knockoff}. As \sys protects a limited ratio of synapses, we use the metric
of the accuracy of protected classes after re-training for 5 epochs. A lower accuracy stands for better protection. 
We compare with three baselines: \qiu, \synapse, and \random. \qiu and \synapse are the same methods used in Section~\ref{section:prune}. \random is to randomly select synapses. 

Figure~\ref{figure:protect} shows the accuracy of the protected classes (Target classes, the left figure) and the accuracy of all classes (All classes, the right figure). It can be observed that, after guarded by \sys, the retrained accuracy on target classes is below 10\%, even when the budget (\ie number of samples) achieves 50k. The small accuracy stands for strong protection. On the contrary, the accuracy of other methods is all above 30\%. For \qiu, the accuracy achieves 90\%, which means it can not protect target classes at all. 

The right figure of Figure~\ref{figure:protect} illustrates why \sys achieves better protection. Compared to \synapse, although the accuracy of \sys on target classes is obviously lower, the accuracy over the whole dataset is higher. It means the accuracy of non-target classes is very high and \sys do not protect them. This trade-off between target classes and non-target classes is similar to the finding in Section~\ref{section:prune} and may be valuable for applications that desire to protect a small set of target classes.

\section{Limitations and discussion}

This section highlights some of the limitations of \sys and discusses possible future directions.

\textbf{DNN architectures.} We only considered five common operations that are commonly used in CNN models, while some operations used in other architectures are not included, such as recurrent neural networks (RNNs) and graph convolutional networks (GCN). These architectures should be easy to support in the future by adding backtracking rules for new operators.

\textbf{Scalability.} In this paper, we did not conduct experiments on very large models and datasets due to limited time. For large DNN models with millions of weights, \sys takes about 10 minutes to compute the slice for an input sample (as shown in Table~\ref{table:evaluation}). Building an adversarial defense (as in Section~\ref{section:defense}) for such a large model may take several days on a single machine. Although the process is slow, especially for in-lab experiments, we think it is acceptable in practice considering the fact that companies usually train a model on large clusters for several weeks.


\textbf{Slicing criterion.} We mainly discuss the slicing criterion concerning only output neurons, but slicing for an intermediate neuron may also be interesting (similar to inspecting an intermediate variable in traditional programs). Such a flexible criterion definition may enable new applications, \eg interpreting or debugging the neural network in finer granularity.

\textbf{More applications.} Beside the three applications discussed in this paper, there are many other applications that are interesting to consider. For example, is it possible to compose different slices to a new model? If it is the case, the way of training networks might be changed. Besides, is it possible to slice certain attributes from a trained model, such as a discriminatory attribute (race, gender, etc.) which we want to exclude from consideration when making decisions?
Last but not least, how can \sys be used to debug model and diagnose fragile weights? Section~\ref{section:defense} has proved its ability to detect adversarial examples, a step forward is to find the deviant neurons or synapses that are critical for errors. Masking them out or adjusting their value may improve the model accuracy.

\textbf{Other slicing techniques.} \sys relies on a set of inputs to compute the slice (\ie dynamic slicing). There are various other slicing techniques that may be interesting to be applied to neural networks. For example, static slicing might be used to compute input-independent slices (as in Section~\ref{section:prune}) much faster as each input doesn't need to be processed separately. Conditioned slicing \cite{canfora1998conditioned} may help the developers to understand the conditions (\eg illumination, viewpoint, etc.) under which the DNN is more vulnerable.
Amorphous slicing may be used to merge neurons and synapses inside the network and slim the network structure~\cite{harman2003amorphous}.



\section{Concluding remarks}

This paper proposes the idea of dynamic slicing on deep neural networks and implements a tool named \sys to compute slices for convolutional neural networks. The working process of \sys consists of a profiling phase, a forward analysis phase, and a backward analysis phase. The profiling and forward analysis phases model the reaction of each neuron based on its activation values. The backward phase traces the data flow recursively from back to front and computes the contributions of each neuron and synapse, which are used to calculate the slice. The usefulness and effectiveness of \sys are demonstrated with three applications on adversarial input detection, targeted model pruning, and selective model protection. The code and data of \sys and all applications will be made available to the community.

\begin{acks}
    We would like to thank the anonymous ESEC/FSE reviewers for their valuable feedback of this paper. We thank Yuxian Qiu and Tribhuvanesh Orekondy for sharing their code. 
    This work was partly supported by the National Key Research and Development Program (2017YFB1001904) and the National Natural Science Foundation of China (61772042).
\end{acks}
\balance
\bibliographystyle{ACM-Reference-Format}
\bibliography{citations}


\begin{thebibliography}{81}


\ifx \showCODEN    \undefined \def \showCODEN     #1{\unskip}     \fi
\ifx \showDOI      \undefined \def \showDOI       #1{#1}\fi
\ifx \showISBNx    \undefined \def \showISBNx     #1{\unskip}     \fi
\ifx \showISBNxiii \undefined \def \showISBNxiii  #1{\unskip}     \fi
\ifx \showISSN     \undefined \def \showISSN      #1{\unskip}     \fi
\ifx \showLCCN     \undefined \def \showLCCN      #1{\unskip}     \fi
\ifx \shownote     \undefined \def \shownote      #1{#1}          \fi
\ifx \showarticletitle \undefined \def \showarticletitle #1{#1}   \fi
\ifx \showURL      \undefined \def \showURL       {\relax}        \fi
\providecommand\bibfield[2]{#2}
\providecommand\bibinfo[2]{#2}
\providecommand\natexlab[1]{#1}
\providecommand\showeprint[2][]{arXiv:#2}

\bibitem[\protect\citeauthoryear{Agrawal, DeMillo, and Spafford}{Agrawal
  et~al\mbox{.}}{1993}]%
        {agrawal1993debugging}
\bibfield{author}{\bibinfo{person}{Hiralal Agrawal}, \bibinfo{person}{Richard~A
  DeMillo}, {and} \bibinfo{person}{Eugene~H Spafford}.}
  \bibinfo{year}{1993}\natexlab{}.
\newblock \showarticletitle{Debugging with dynamic slicing and backtracking}.
\newblock \bibinfo{journal}{\emph{Software: Practice and Experience}}
  \bibinfo{volume}{23}, \bibinfo{number}{6} (\bibinfo{year}{1993}),
  \bibinfo{pages}{589--616}.
\newblock


\bibitem[\protect\citeauthoryear{Alaifari, Alberti, and Gauksson}{Alaifari
  et~al\mbox{.}}{2018}]%
        {alaifari2018adef}
\bibfield{author}{\bibinfo{person}{Rima Alaifari}, \bibinfo{person}{Giovanni~S
  Alberti}, {and} \bibinfo{person}{Tandri Gauksson}.}
  \bibinfo{year}{2018}\natexlab{}.
\newblock \showarticletitle{ADef: an iterative algorithm to construct
  adversarial deformations}.
\newblock \bibinfo{journal}{\emph{arXiv preprint arXiv:1804.07729}}
  (\bibinfo{year}{2018}).
\newblock


\bibitem[\protect\citeauthoryear{Arlt, Podelski, and Wehrle}{Arlt
  et~al\mbox{.}}{2014}]%
        {arlt2014reducing}
\bibfield{author}{\bibinfo{person}{Stephan Arlt}, \bibinfo{person}{Andreas
  Podelski}, {and} \bibinfo{person}{Martin Wehrle}.}
  \bibinfo{year}{2014}\natexlab{}.
\newblock \showarticletitle{Reducing GUI test suites via program slicing}. In
  \bibinfo{booktitle}{\emph{Proceedings of the 2014 International Symposium on
  Software Testing and Analysis}}. ACM, \bibinfo{pages}{270--281}.
\newblock


\bibitem[\protect\citeauthoryear{Azim, Alavi, Neamtiu, and Gupta}{Azim
  et~al\mbox{.}}{2019}]%
        {azim2019dynamic}
\bibfield{author}{\bibinfo{person}{Tanzirul Azim}, \bibinfo{person}{Arash
  Alavi}, \bibinfo{person}{Iulian Neamtiu}, {and} \bibinfo{person}{Rajiv
  Gupta}.} \bibinfo{year}{2019}\natexlab{}.
\newblock \showarticletitle{Dynamic slicing for android}. In
  \bibinfo{booktitle}{\emph{2019 IEEE/ACM 41st International Conference on
  Software Engineering (ICSE)}}. IEEE, \bibinfo{pages}{1154--1164}.
\newblock


\bibitem[\protect\citeauthoryear{Binkley, Gold, Harman, Islam, Krinke, and
  Yoo}{Binkley et~al\mbox{.}}{2014}]%
        {binkley2014orbs}
\bibfield{author}{\bibinfo{person}{David Binkley}, \bibinfo{person}{Nicolas
  Gold}, \bibinfo{person}{Mark Harman}, \bibinfo{person}{Syed Islam},
  \bibinfo{person}{Jens Krinke}, {and} \bibinfo{person}{Shin Yoo}.}
  \bibinfo{year}{2014}\natexlab{}.
\newblock \showarticletitle{ORBS: Language-independent program slicing}. In
  \bibinfo{booktitle}{\emph{22nd ACM SIGSOFT International Symposium on
  Foundations of Software Engineering}}. ACM, \bibinfo{pages}{109--120}.
\newblock


\bibitem[\protect\citeauthoryear{Binkley and Harman}{Binkley and
  Harman}{2004}]%
        {binkley2004survey}
\bibfield{author}{\bibinfo{person}{David~W Binkley} {and} \bibinfo{person}{Mark
  Harman}.} \bibinfo{year}{2004}\natexlab{}.
\newblock \showarticletitle{A survey of empirical results on program slicing.}
\newblock \bibinfo{journal}{\emph{Adv. Comput.}} \bibinfo{volume}{62},
  \bibinfo{number}{105178} (\bibinfo{year}{2004}), \bibinfo{pages}{105--178}.
\newblock


\bibitem[\protect\citeauthoryear{Breiman}{Breiman}{2017}]%
        {breiman2017classification}
\bibfield{author}{\bibinfo{person}{Leo Breiman}.}
  \bibinfo{year}{2017}\natexlab{}.
\newblock \bibinfo{booktitle}{\emph{Classification and regression trees}}.
\newblock \bibinfo{publisher}{Routledge}.
\newblock


\bibitem[\protect\citeauthoryear{Brendel, Rauber, and Bethge}{Brendel
  et~al\mbox{.}}{2017}]%
        {brendel2017decision}
\bibfield{author}{\bibinfo{person}{Wieland Brendel}, \bibinfo{person}{Jonas
  Rauber}, {and} \bibinfo{person}{Matthias Bethge}.}
  \bibinfo{year}{2017}\natexlab{}.
\newblock \showarticletitle{Decision-based adversarial attacks: Reliable
  attacks against black-box machine learning models}.
\newblock \bibinfo{journal}{\emph{arXiv preprint arXiv:1712.04248}}
  (\bibinfo{year}{2017}).
\newblock


\bibitem[\protect\citeauthoryear{Cai, Chen, Ooi, and Gao}{Cai
  et~al\mbox{.}}{2019}]%
        {cai2019model}
\bibfield{author}{\bibinfo{person}{Shaofeng Cai}, \bibinfo{person}{Gang Chen},
  \bibinfo{person}{Beng~Chin Ooi}, {and} \bibinfo{person}{Jinyang Gao}.}
  \bibinfo{year}{2019}\natexlab{}.
\newblock \showarticletitle{Model slicing for supporting complex analytics with
  elastic inference cost and resource constraints}.
\newblock \bibinfo{journal}{\emph{VLDB Endowment}} \bibinfo{volume}{13},
  \bibinfo{number}{2} (\bibinfo{year}{2019}), \bibinfo{pages}{86--99}.
\newblock


\bibitem[\protect\citeauthoryear{Canfora, Cimitile, and De~Lucia}{Canfora
  et~al\mbox{.}}{1998}]%
        {canfora1998conditioned}
\bibfield{author}{\bibinfo{person}{Gerardo Canfora}, \bibinfo{person}{Aniello
  Cimitile}, {and} \bibinfo{person}{Andrea De~Lucia}.}
  \bibinfo{year}{1998}\natexlab{}.
\newblock \showarticletitle{Conditioned program slicing}.
\newblock \bibinfo{journal}{\emph{Information and Software Technology}}
  \bibinfo{volume}{40}, \bibinfo{number}{11-12} (\bibinfo{year}{1998}),
  \bibinfo{pages}{595--607}.
\newblock


\bibitem[\protect\citeauthoryear{Carlini and Wagner}{Carlini and
  Wagner}{2017}]%
        {carlini2017towards}
\bibfield{author}{\bibinfo{person}{Nicholas Carlini} {and}
  \bibinfo{person}{David Wagner}.} \bibinfo{year}{2017}\natexlab{}.
\newblock \showarticletitle{Towards evaluating the robustness of neural
  networks}. In \bibinfo{booktitle}{\emph{2017 {IEEE} Symposium on Security and
  Privacy ({SP})}}. IEEE, \bibinfo{pages}{39--57}.
\newblock


\bibitem[\protect\citeauthoryear{Chou, Beal, Levy, Yeung, Haque, and
  Fei-Fei}{Chou et~al\mbox{.}}{2018}]%
        {chou2018faster}
\bibfield{author}{\bibinfo{person}{Edward Chou}, \bibinfo{person}{Josh Beal},
  \bibinfo{person}{Daniel Levy}, \bibinfo{person}{Serena Yeung},
  \bibinfo{person}{Albert Haque}, {and} \bibinfo{person}{Li Fei-Fei}.}
  \bibinfo{year}{2018}\natexlab{}.
\newblock \showarticletitle{Faster cryptonets: Leveraging sparsity for
  real-world encrypted inference}.
\newblock \bibinfo{journal}{\emph{arXiv preprint arXiv:1811.09953}}
  (\bibinfo{year}{2018}).
\newblock


\bibitem[\protect\citeauthoryear{Cire{\c{s}}An, Meier, Masci, and
  Schmidhuber}{Cire{\c{s}}An et~al\mbox{.}}{2012}]%
        {cirecsan2012multi}
\bibfield{author}{\bibinfo{person}{Dan Cire{\c{s}}An}, \bibinfo{person}{Ueli
  Meier}, \bibinfo{person}{Jonathan Masci}, {and} \bibinfo{person}{J{\"u}rgen
  Schmidhuber}.} \bibinfo{year}{2012}\natexlab{}.
\newblock \showarticletitle{Multi-column deep neural network for traffic sign
  classification}.
\newblock \bibinfo{journal}{\emph{Neural networks}}  \bibinfo{volume}{32}
  (\bibinfo{year}{2012}), \bibinfo{pages}{333--338}.
\newblock


\bibitem[\protect\citeauthoryear{Clarke, Fujita, Rajan, Reps, Shankar, and
  Teitelbaum}{Clarke et~al\mbox{.}}{1999}]%
        {clarke1999program}
\bibfield{author}{\bibinfo{person}{Edmund~M Clarke}, \bibinfo{person}{Masahiro
  Fujita}, \bibinfo{person}{Sreeranga~P Rajan}, \bibinfo{person}{T Reps},
  \bibinfo{person}{Subash Shankar}, {and} \bibinfo{person}{Tim Teitelbaum}.}
  \bibinfo{year}{1999}\natexlab{}.
\newblock \showarticletitle{Program slicing of hardware description languages}.
  In \bibinfo{booktitle}{\emph{Advanced Research Working Conference on Correct
  Hardware Design and Verification Methods}}. Springer,
  \bibinfo{pages}{298--313}.
\newblock


\bibitem[\protect\citeauthoryear{Correia-Silva, Berriel, Badue, de~Souza, and
  Oliveira-Santos}{Correia-Silva et~al\mbox{.}}{2018}]%
        {correia2018copycat}
\bibfield{author}{\bibinfo{person}{Jacson~Rodrigues Correia-Silva},
  \bibinfo{person}{Rodrigo~F Berriel}, \bibinfo{person}{Claudine Badue},
  \bibinfo{person}{Alberto~F de Souza}, {and} \bibinfo{person}{Thiago
  Oliveira-Santos}.} \bibinfo{year}{2018}\natexlab{}.
\newblock \showarticletitle{Copycat CNN: Stealing knowledge by persuading
  confession with random non-labeled data}. In \bibinfo{booktitle}{\emph{2018
  International Joint Conference on Neural Networks (IJCNN)}}. IEEE,
  \bibinfo{pages}{1--8}.
\newblock


\bibitem[\protect\citeauthoryear{Costan and Devadas}{Costan and
  Devadas}{2016}]%
        {costan2016intel}
\bibfield{author}{\bibinfo{person}{Victor Costan} {and}
  \bibinfo{person}{Srinivas Devadas}.} \bibinfo{year}{2016}\natexlab{}.
\newblock \showarticletitle{Intel SGX Explained.}
\newblock \bibinfo{journal}{\emph{IACR Cryptology ePrint Archive}}
  \bibinfo{volume}{2016}, \bibinfo{number}{086} (\bibinfo{year}{2016}),
  \bibinfo{pages}{1--118}.
\newblock


\bibitem[\protect\citeauthoryear{Costan, Lebedev, and Devadas}{Costan
  et~al\mbox{.}}{2016}]%
        {costan2016sanctum}
\bibfield{author}{\bibinfo{person}{Victor Costan}, \bibinfo{person}{Ilia
  Lebedev}, {and} \bibinfo{person}{Srinivas Devadas}.}
  \bibinfo{year}{2016}\natexlab{}.
\newblock \showarticletitle{Sanctum: Minimal hardware extensions for strong
  software isolation}. In \bibinfo{booktitle}{\emph{25th {USENIX} Security
  Symposium ({USENIX} Security 16)}}. \bibinfo{pages}{857--874}.
\newblock


\bibitem[\protect\citeauthoryear{Devlin, Chang, Lee, and Toutanova}{Devlin
  et~al\mbox{.}}{2019}]%
        {devlin2018bert}
\bibfield{author}{\bibinfo{person}{Jacob Devlin}, \bibinfo{person}{Ming-Wei
  Chang}, \bibinfo{person}{Kenton Lee}, {and} \bibinfo{person}{Kristina
  Toutanova}.} \bibinfo{year}{2019}\natexlab{}.
\newblock \showarticletitle{{BERT}: Pre-training of Deep Bidirectional
  Transformers for Language Understanding}.
\newblock  (\bibinfo{year}{2019}).
\newblock


\bibitem[\protect\citeauthoryear{Du, Xie, Li, Ma, Liu, and Zhao}{Du
  et~al\mbox{.}}{2019}]%
        {du2019deepstellar}
\bibfield{author}{\bibinfo{person}{Xiaoning Du}, \bibinfo{person}{Xiaofei Xie},
  \bibinfo{person}{Yi Li}, \bibinfo{person}{Lei Ma}, \bibinfo{person}{Yang
  Liu}, {and} \bibinfo{person}{Jianjun Zhao}.} \bibinfo{year}{2019}\natexlab{}.
\newblock \showarticletitle{Deepstellar: model-based quantitative analysis of
  stateful deep learning systems}. In \bibinfo{booktitle}{\emph{27th ACM Joint
  Meeting on European Software Engineering Conference and Symposium on the
  Foundations of Software Engineering}}. \bibinfo{pages}{477--487}.
\newblock


\bibitem[\protect\citeauthoryear{Elkahky, Song, and He}{Elkahky
  et~al\mbox{.}}{2015}]%
        {elkahky2015multi}
\bibfield{author}{\bibinfo{person}{Ali~Mamdouh Elkahky}, \bibinfo{person}{Yang
  Song}, {and} \bibinfo{person}{Xiaodong He}.} \bibinfo{year}{2015}\natexlab{}.
\newblock \showarticletitle{A multi-view deep learning approach for cross
  domain user modeling in recommendation systems}. In
  \bibinfo{booktitle}{\emph{24th International Conference on World Wide Web}}.
  \bibinfo{pages}{278--288}.
\newblock


\bibitem[\protect\citeauthoryear{Engstrom, Tsipras, Schmidt, and
  Madry}{Engstrom et~al\mbox{.}}{2017}]%
        {engstrom2017rotation}
\bibfield{author}{\bibinfo{person}{Logan Engstrom}, \bibinfo{person}{Dimitris
  Tsipras}, \bibinfo{person}{Ludwig Schmidt}, {and} \bibinfo{person}{Aleksander
  Madry}.} \bibinfo{year}{2017}\natexlab{}.
\newblock \showarticletitle{A rotation and a translation suffice: Fooling cnns
  with simple transformations}.
\newblock \bibinfo{journal}{\emph{arXiv preprint arXiv:1712.02779}}
  \bibinfo{volume}{1}, \bibinfo{number}{2} (\bibinfo{year}{2017}),
  \bibinfo{pages}{3}.
\newblock


\bibitem[\protect\citeauthoryear{Fidel, Bitton, and Shabtai}{Fidel
  et~al\mbox{.}}{2019}]%
        {fidel2019explainability}
\bibfield{author}{\bibinfo{person}{Gil Fidel}, \bibinfo{person}{Ron Bitton},
  {and} \bibinfo{person}{Asaf Shabtai}.} \bibinfo{year}{2019}\natexlab{}.
\newblock \showarticletitle{When Explainability Meets Adversarial Learning:
  Detecting Adversarial Examples using SHAP Signatures}.
\newblock \bibinfo{journal}{\emph{arXiv preprint arXiv:1909.03418}}
  (\bibinfo{year}{2019}).
\newblock


\bibitem[\protect\citeauthoryear{Gallagher and Lyle}{Gallagher and
  Lyle}{1991}]%
        {gallagher1991using}
\bibfield{author}{\bibinfo{person}{Keith~Brian Gallagher} {and}
  \bibinfo{person}{James~R Lyle}.} \bibinfo{year}{1991}\natexlab{}.
\newblock \showarticletitle{Using program slicing in software maintenance}.
\newblock \bibinfo{journal}{\emph{IEEE transactions on software engineering}}
  \bibinfo{number}{8} (\bibinfo{year}{1991}), \bibinfo{pages}{751--761}.
\newblock


\bibitem[\protect\citeauthoryear{Gehr, Mirman, Drachsler-Cohen, Tsankov,
  Chaudhuri, and Vechev}{Gehr et~al\mbox{.}}{2018}]%
        {gehr2018ai2}
\bibfield{author}{\bibinfo{person}{Timon Gehr}, \bibinfo{person}{Matthew
  Mirman}, \bibinfo{person}{Dana Drachsler-Cohen}, \bibinfo{person}{Petar
  Tsankov}, \bibinfo{person}{Swarat Chaudhuri}, {and} \bibinfo{person}{Martin
  Vechev}.} \bibinfo{year}{2018}\natexlab{}.
\newblock \showarticletitle{{AI2}: Safety and robustness certification of
  neural networks with abstract interpretation}. In
  \bibinfo{booktitle}{\emph{2018 IEEE Symposium on Security and Privacy (SP)}}.
  IEEE, \bibinfo{pages}{3--18}.
\newblock


\bibitem[\protect\citeauthoryear{Gers and Schmidhuber}{Gers and
  Schmidhuber}{2001}]%
        {gers2001lstm}
\bibfield{author}{\bibinfo{person}{Felix~A Gers} {and} \bibinfo{person}{E
  Schmidhuber}.} \bibinfo{year}{2001}\natexlab{}.
\newblock \showarticletitle{LSTM recurrent networks learn simple context-free
  and context-sensitive languages}.
\newblock \bibinfo{journal}{\emph{IEEE Transactions on Neural Networks}}
  \bibinfo{volume}{12}, \bibinfo{number}{6} (\bibinfo{year}{2001}),
  \bibinfo{pages}{1333--1340}.
\newblock


\bibitem[\protect\citeauthoryear{Gilad-Bachrach, Dowlin, Laine, Lauter,
  Naehrig, and Wernsing}{Gilad-Bachrach et~al\mbox{.}}{2016}]%
        {gilad2016cryptonets}
\bibfield{author}{\bibinfo{person}{Ran Gilad-Bachrach}, \bibinfo{person}{Nathan
  Dowlin}, \bibinfo{person}{Kim Laine}, \bibinfo{person}{Kristin Lauter},
  \bibinfo{person}{Michael Naehrig}, {and} \bibinfo{person}{John Wernsing}.}
  \bibinfo{year}{2016}\natexlab{}.
\newblock \showarticletitle{Cryptonets: Applying neural networks to encrypted
  data with high throughput and accuracy}. In
  \bibinfo{booktitle}{\emph{International Conference on Machine Learning}}.
  \bibinfo{pages}{201--210}.
\newblock


\bibitem[\protect\citeauthoryear{Goodfellow, Shlens, and Szegedy}{Goodfellow
  et~al\mbox{.}}{2014}]%
        {goodfellow2014explaining}
\bibfield{author}{\bibinfo{person}{Ian~J Goodfellow}, \bibinfo{person}{Jonathon
  Shlens}, {and} \bibinfo{person}{Christian Szegedy}.}
  \bibinfo{year}{2014}\natexlab{}.
\newblock \showarticletitle{Explaining and harnessing adversarial examples}.
\newblock \bibinfo{journal}{\emph{arXiv preprint arXiv:1412.6572}}
  (\bibinfo{year}{2014}).
\newblock


\bibitem[\protect\citeauthoryear{Gopinath, Converse, Pasareanu, and
  Taly}{Gopinath et~al\mbox{.}}{2019}]%
        {gopinath2019property}
\bibfield{author}{\bibinfo{person}{Divya Gopinath}, \bibinfo{person}{Hayes
  Converse}, \bibinfo{person}{Corina Pasareanu}, {and} \bibinfo{person}{Ankur
  Taly}.} \bibinfo{year}{2019}\natexlab{}.
\newblock \showarticletitle{Property inference for deep neural networks}. In
  \bibinfo{booktitle}{\emph{34th IEEE/ACM International Conference on Automated
  Software Engineering (ASE)}}. IEEE, \bibinfo{pages}{797--809}.
\newblock


\bibitem[\protect\citeauthoryear{Gopinath, Wang, Zhang, Pasareanu, and
  Khurshid}{Gopinath et~al\mbox{.}}{2018}]%
        {gopinath2018symbolic}
\bibfield{author}{\bibinfo{person}{Divya Gopinath}, \bibinfo{person}{Kaiyuan
  Wang}, \bibinfo{person}{Mengshi Zhang}, \bibinfo{person}{Corina~S Pasareanu},
  {and} \bibinfo{person}{Sarfraz Khurshid}.} \bibinfo{year}{2018}\natexlab{}.
\newblock \showarticletitle{Symbolic execution for deep neural networks}.
\newblock \bibinfo{journal}{\emph{arXiv preprint arXiv:1807.10439}}
  (\bibinfo{year}{2018}).
\newblock


\bibitem[\protect\citeauthoryear{Gu, Zhang, Zhang, and Kim}{Gu
  et~al\mbox{.}}{2016}]%
        {gu2016deep_api}
\bibfield{author}{\bibinfo{person}{Xiaodong Gu}, \bibinfo{person}{Hongyu
  Zhang}, \bibinfo{person}{Dongmei Zhang}, {and} \bibinfo{person}{Sunghun
  Kim}.} \bibinfo{year}{2016}\natexlab{}.
\newblock \showarticletitle{Deep API learning}. In
  \bibinfo{booktitle}{\emph{24th ACM SIGSOFT International Symposium on
  Foundations of Software Engineering}}. \bibinfo{pages}{631--642}.
\newblock


\bibitem[\protect\citeauthoryear{Han, Mao, and Dally}{Han
  et~al\mbox{.}}{2016}]%
        {han2015deep_compression}
\bibfield{author}{\bibinfo{person}{Song Han}, \bibinfo{person}{Huizi Mao},
  {and} \bibinfo{person}{William~J Dally}.} \bibinfo{year}{2016}\natexlab{}.
\newblock \showarticletitle{Deep compression: compressing deep neural networks
  with pruning, trained quantization and huffman coding}. In
  \bibinfo{booktitle}{\emph{4th International Conference on Learning
  Representations, {ICLR}}}.
\newblock


\bibitem[\protect\citeauthoryear{Han, Pool, Tran, and Dally}{Han
  et~al\mbox{.}}{2015}]%
        {han2015pruning}
\bibfield{author}{\bibinfo{person}{Song Han}, \bibinfo{person}{Jeff Pool},
  \bibinfo{person}{John Tran}, {and} \bibinfo{person}{William Dally}.}
  \bibinfo{year}{2015}\natexlab{}.
\newblock \showarticletitle{Learning both weights and connections for efficient
  neural network}. In \bibinfo{booktitle}{\emph{Advances in neural information
  processing systems}}. \bibinfo{pages}{1135--1143}.
\newblock


\bibitem[\protect\citeauthoryear{Harman, Binkley, and Danicic}{Harman
  et~al\mbox{.}}{2003}]%
        {harman2003amorphous}
\bibfield{author}{\bibinfo{person}{Mark Harman}, \bibinfo{person}{David
  Binkley}, {and} \bibinfo{person}{Sebastian Danicic}.}
  \bibinfo{year}{2003}\natexlab{}.
\newblock \showarticletitle{Amorphous program slicing}.
\newblock \bibinfo{journal}{\emph{Journal of Systems and Software}}
  \bibinfo{volume}{68}, \bibinfo{number}{1} (\bibinfo{year}{2003}),
  \bibinfo{pages}{45--64}.
\newblock


\bibitem[\protect\citeauthoryear{He, Zhang, Ren, and Sun}{He
  et~al\mbox{.}}{2016}]%
        {he2016deep}
\bibfield{author}{\bibinfo{person}{Kaiming He}, \bibinfo{person}{Xiangyu
  Zhang}, \bibinfo{person}{Shaoqing Ren}, {and} \bibinfo{person}{Jian Sun}.}
  \bibinfo{year}{2016}\natexlab{}.
\newblock \showarticletitle{Deep residual learning for image recognition}. In
  \bibinfo{booktitle}{\emph{IEEE conference on computer vision and pattern
  recognition ({CVPR})}}. \bibinfo{pages}{770--778}.
\newblock


\bibitem[\protect\citeauthoryear{He, Zhang, and Sun}{He et~al\mbox{.}}{2017}]%
        {he2017channel}
\bibfield{author}{\bibinfo{person}{Yihui He}, \bibinfo{person}{Xiangyu Zhang},
  {and} \bibinfo{person}{Jian Sun}.} \bibinfo{year}{2017}\natexlab{}.
\newblock \showarticletitle{Channel pruning for accelerating very deep neural
  networks}. In \bibinfo{booktitle}{\emph{IEEE International Conference on
  Computer Vision ({CVPR})}}. \bibinfo{pages}{1389--1397}.
\newblock


\bibitem[\protect\citeauthoryear{Horwitz, Reps, and Binkley}{Horwitz
  et~al\mbox{.}}{1990}]%
        {horwitz1990interprocedural}
\bibfield{author}{\bibinfo{person}{Susan Horwitz}, \bibinfo{person}{Thomas
  Reps}, {and} \bibinfo{person}{David Binkley}.}
  \bibinfo{year}{1990}\natexlab{}.
\newblock \showarticletitle{Interprocedural slicing using dependence graphs}.
\newblock \bibinfo{journal}{\emph{ACM Transactions on Programming Languages and
  Systems (TOPLAS)}} \bibinfo{volume}{12}, \bibinfo{number}{1}
  (\bibinfo{year}{1990}), \bibinfo{pages}{26--60}.
\newblock


\bibitem[\protect\citeauthoryear{Hu, Liang, Deng, Li, Xie, Ji, Ding, Liu,
  Sherwood, and Xie}{Hu et~al\mbox{.}}{2019}]%
        {hu2019neural}
\bibfield{author}{\bibinfo{person}{Xing Hu}, \bibinfo{person}{Ling Liang},
  \bibinfo{person}{Lei Deng}, \bibinfo{person}{Shuangchen Li},
  \bibinfo{person}{Xinfeng Xie}, \bibinfo{person}{Yu Ji},
  \bibinfo{person}{Yufei Ding}, \bibinfo{person}{Chang Liu},
  \bibinfo{person}{Timothy Sherwood}, {and} \bibinfo{person}{Yuan Xie}.}
  \bibinfo{year}{2019}\natexlab{}.
\newblock \showarticletitle{Neural network model extraction attacks in edge
  devices by hearing architectural hints}.
\newblock \bibinfo{journal}{\emph{arXiv preprint arXiv:1903.03916}}
  (\bibinfo{year}{2019}).
\newblock


\bibitem[\protect\citeauthoryear{Jagielski, Carlini, Berthelot, Kurakin, and
  Papernot}{Jagielski et~al\mbox{.}}{2019}]%
        {jagielski2019high}
\bibfield{author}{\bibinfo{person}{Matthew Jagielski},
  \bibinfo{person}{Nicholas Carlini}, \bibinfo{person}{David Berthelot},
  \bibinfo{person}{Alex Kurakin}, {and} \bibinfo{person}{Nicolas Papernot}.}
  \bibinfo{year}{2019}\natexlab{}.
\newblock \showarticletitle{High-fidelity extraction of neural network models}.
\newblock \bibinfo{journal}{\emph{arXiv preprint arXiv:1909.01838}}
  (\bibinfo{year}{2019}).
\newblock


\bibitem[\protect\citeauthoryear{Juuti, Szyller, Marchal, and Asokan}{Juuti
  et~al\mbox{.}}{2019}]%
        {juuti2019prada}
\bibfield{author}{\bibinfo{person}{Mika Juuti}, \bibinfo{person}{Sebastian
  Szyller}, \bibinfo{person}{Samuel Marchal}, {and} \bibinfo{person}{N
  Asokan}.} \bibinfo{year}{2019}\natexlab{}.
\newblock \showarticletitle{PRADA: protecting against DNN model stealing
  attacks}. In \bibinfo{booktitle}{\emph{IEEE European Symposium on Security
  and Privacy (EuroS\&P)}}. \bibinfo{pages}{512--527}.
\newblock


\bibitem[\protect\citeauthoryear{Katz, Barrett, Dill, Julian, and
  Kochenderfer}{Katz et~al\mbox{.}}{2017}]%
        {katz2017reluplex}
\bibfield{author}{\bibinfo{person}{Guy Katz}, \bibinfo{person}{Clark Barrett},
  \bibinfo{person}{David~L Dill}, \bibinfo{person}{Kyle Julian}, {and}
  \bibinfo{person}{Mykel~J Kochenderfer}.} \bibinfo{year}{2017}\natexlab{}.
\newblock \showarticletitle{Reluplex: An efficient SMT solver for verifying
  deep neural networks}. In \bibinfo{booktitle}{\emph{International Conference
  on Computer Aided Verification}}. Springer, \bibinfo{pages}{97--117}.
\newblock


\bibitem[\protect\citeauthoryear{Kocher, Jaffe, and Jun}{Kocher
  et~al\mbox{.}}{1999}]%
        {kocher1999differential}
\bibfield{author}{\bibinfo{person}{Paul Kocher}, \bibinfo{person}{Joshua
  Jaffe}, {and} \bibinfo{person}{Benjamin Jun}.}
  \bibinfo{year}{1999}\natexlab{}.
\newblock \showarticletitle{Differential power analysis}. In
  \bibinfo{booktitle}{\emph{Annual International Cryptology Conference}}.
  Springer, \bibinfo{pages}{388--397}.
\newblock


\bibitem[\protect\citeauthoryear{Korel and Laski}{Korel and Laski}{1988}]%
        {korel1988dynamic}
\bibfield{author}{\bibinfo{person}{Bogdan Korel} {and} \bibinfo{person}{Janusz
  Laski}.} \bibinfo{year}{1988}\natexlab{}.
\newblock \showarticletitle{Dynamic program slicing}.
\newblock \bibinfo{journal}{\emph{Inform. Process. Lett.}}
  \bibinfo{volume}{29}, \bibinfo{number}{3} (\bibinfo{year}{1988}),
  \bibinfo{pages}{155--163}.
\newblock


\bibitem[\protect\citeauthoryear{Li, Bai, Zhou, Xie, Zhang, and Yuille}{Li
  et~al\mbox{.}}{2020}]%
        {li2020learning}
\bibfield{author}{\bibinfo{person}{Yingwei Li}, \bibinfo{person}{Song Bai},
  \bibinfo{person}{Yuyin Zhou}, \bibinfo{person}{Cihang Xie},
  \bibinfo{person}{Zhishuai Zhang}, {and} \bibinfo{person}{Alan Yuille}.}
  \bibinfo{year}{2020}\natexlab{}.
\newblock \showarticletitle{Learning Transferable Adversarial Examples via
  Ghost Networks}. In \bibinfo{booktitle}{\emph{AAAI Conference on Artificial
  Intelligence}}, Vol.~\bibinfo{volume}{34}.
\newblock


\bibitem[\protect\citeauthoryear{Li, Chen, Li, Guo, Huang, Fredrikson, Agarwal,
  and Hong}{Li et~al\mbox{.}}{2017}]%
        {li2017privacystreams}
\bibfield{author}{\bibinfo{person}{Yuanchun Li}, \bibinfo{person}{Fanglin
  Chen}, \bibinfo{person}{Toby Jia-Jun Li}, \bibinfo{person}{Yao Guo},
  \bibinfo{person}{Gang Huang}, \bibinfo{person}{Matthew Fredrikson},
  \bibinfo{person}{Yuvraj Agarwal}, {and} \bibinfo{person}{Jason~I Hong}.}
  \bibinfo{year}{2017}\natexlab{}.
\newblock \showarticletitle{Privacystreams: Enabling transparency in personal
  data processing for mobile apps}.
\newblock \bibinfo{journal}{\emph{ACM on Interactive, Mobile, Wearable and
  Ubiquitous Technologies}} \bibinfo{volume}{1}, \bibinfo{number}{3}
  (\bibinfo{year}{2017}), \bibinfo{pages}{1--26}.
\newblock


\bibitem[\protect\citeauthoryear{Li, Yang, Guo, and Chen}{Li
  et~al\mbox{.}}{2019}]%
        {li2019humanoid}
\bibfield{author}{\bibinfo{person}{Yuanchun Li}, \bibinfo{person}{Ziyue Yang},
  \bibinfo{person}{Yao Guo}, {and} \bibinfo{person}{Xiangqun Chen}.}
  \bibinfo{year}{2019}\natexlab{}.
\newblock \showarticletitle{Humanoid: a deep learning-based approach to
  automated black-box Android app testing}. In \bibinfo{booktitle}{\emph{2019
  34th IEEE/ACM International Conference on Automated Software Engineering
  (ASE)}}. IEEE, \bibinfo{pages}{1070--1073}.
\newblock


\bibitem[\protect\citeauthoryear{Liu, Sun, Zhou, Huang, and Darrell}{Liu
  et~al\mbox{.}}{2018}]%
        {liu2018rethinking}
\bibfield{author}{\bibinfo{person}{Zhuang Liu}, \bibinfo{person}{Mingjie Sun},
  \bibinfo{person}{Tinghui Zhou}, \bibinfo{person}{Gao Huang}, {and}
  \bibinfo{person}{Trevor Darrell}.} \bibinfo{year}{2018}\natexlab{}.
\newblock \showarticletitle{Rethinking the value of network pruning}.
\newblock \bibinfo{journal}{\emph{arXiv preprint arXiv:1810.05270}}
  (\bibinfo{year}{2018}).
\newblock


\bibitem[\protect\citeauthoryear{Ma and Liu}{Ma and Liu}{2019}]%
        {ma2019nic}
\bibfield{author}{\bibinfo{person}{Shiqing Ma} {and} \bibinfo{person}{Yingqi
  Liu}.} \bibinfo{year}{2019}\natexlab{}.
\newblock \showarticletitle{{NIC}: Detecting adversarial samples with neural
  network invariant checking}. In \bibinfo{booktitle}{\emph{26th Network and
  Distributed System Security Symposium (NDSS)}}.
\newblock


\bibitem[\protect\citeauthoryear{Ma, Li, Wang, Erfani, Wijewickrema,
  Schoenebeck, Song, Houle, and Bailey}{Ma et~al\mbox{.}}{2018}]%
        {ma2018characterizing}
\bibfield{author}{\bibinfo{person}{Xingjun Ma}, \bibinfo{person}{Bo Li},
  \bibinfo{person}{Yisen Wang}, \bibinfo{person}{Sarah~M Erfani},
  \bibinfo{person}{Sudanthi Wijewickrema}, \bibinfo{person}{Grant Schoenebeck},
  \bibinfo{person}{Dawn Song}, \bibinfo{person}{Michael~E Houle}, {and}
  \bibinfo{person}{James Bailey}.} \bibinfo{year}{2018}\natexlab{}.
\newblock \showarticletitle{Characterizing adversarial subspaces using local
  intrinsic dimensionality}.
\newblock \bibinfo{journal}{\emph{arXiv preprint arXiv:1801.02613}}
  (\bibinfo{year}{2018}).
\newblock


\bibitem[\protect\citeauthoryear{Madry, Makelov, Schmidt, Tsipras, and
  Vladu}{Madry et~al\mbox{.}}{2017}]%
        {madry2017towards}
\bibfield{author}{\bibinfo{person}{Aleksander Madry},
  \bibinfo{person}{Aleksandar Makelov}, \bibinfo{person}{Ludwig Schmidt},
  \bibinfo{person}{Dimitris Tsipras}, {and} \bibinfo{person}{Adrian Vladu}.}
  \bibinfo{year}{2017}\natexlab{}.
\newblock \showarticletitle{Towards deep learning models resistant to
  adversarial attacks}.
\newblock \bibinfo{journal}{\emph{arXiv preprint arXiv:1706.06083}}
  (\bibinfo{year}{2017}).
\newblock


\bibitem[\protect\citeauthoryear{Moosavi-Dezfooli, Fawzi, and
  Frossard}{Moosavi-Dezfooli et~al\mbox{.}}{2016}]%
        {moosavi2016deepfool}
\bibfield{author}{\bibinfo{person}{Seyed-Mohsen Moosavi-Dezfooli},
  \bibinfo{person}{Alhussein Fawzi}, {and} \bibinfo{person}{Pascal Frossard}.}
  \bibinfo{year}{2016}\natexlab{}.
\newblock \showarticletitle{Deepfool: a simple and accurate method to fool deep
  neural networks}. In \bibinfo{booktitle}{\emph{IEEE conference on computer
  vision and pattern recognition}}. \bibinfo{pages}{2574--2582}.
\newblock


\bibitem[\protect\citeauthoryear{Mustafa, Khan, Hayat, Goecke, Shen, and
  Shao}{Mustafa et~al\mbox{.}}{2019}]%
        {mustafa2019adversarial}
\bibfield{author}{\bibinfo{person}{Aamir Mustafa}, \bibinfo{person}{Salman
  Khan}, \bibinfo{person}{Munawar Hayat}, \bibinfo{person}{Roland Goecke},
  \bibinfo{person}{Jianbing Shen}, {and} \bibinfo{person}{Ling Shao}.}
  \bibinfo{year}{2019}\natexlab{}.
\newblock \showarticletitle{Adversarial defense by restricting the hidden space
  of deep neural networks}. In \bibinfo{booktitle}{\emph{IEEE International
  Conference on Computer Vision (CVPR)}}. \bibinfo{pages}{3385--3394}.
\newblock


\bibitem[\protect\citeauthoryear{Narodytska and Kasiviswanathan}{Narodytska and
  Kasiviswanathan}{2016}]%
        {narodytska2016simple}
\bibfield{author}{\bibinfo{person}{Nina Narodytska} {and}
  \bibinfo{person}{Shiva~Prasad Kasiviswanathan}.}
  \bibinfo{year}{2016}\natexlab{}.
\newblock \showarticletitle{Simple black-box adversarial perturbations for deep
  networks}.
\newblock \bibinfo{journal}{\emph{arXiv preprint arXiv:1612.06299}}
  (\bibinfo{year}{2016}).
\newblock


\bibitem[\protect\citeauthoryear{Orekondy, Schiele, and Fritz}{Orekondy
  et~al\mbox{.}}{2019a}]%
        {orekondy2019knockoff}
\bibfield{author}{\bibinfo{person}{Tribhuvanesh Orekondy},
  \bibinfo{person}{Bernt Schiele}, {and} \bibinfo{person}{Mario Fritz}.}
  \bibinfo{year}{2019}\natexlab{a}.
\newblock \showarticletitle{Knockoff nets: Stealing functionality of black-box
  models}. In \bibinfo{booktitle}{\emph{IEEE Conference on Computer Vision and
  Pattern Recognition (CVPR)}}. \bibinfo{pages}{4954--4963}.
\newblock


\bibitem[\protect\citeauthoryear{Orekondy, Schiele, and Fritz}{Orekondy
  et~al\mbox{.}}{2019b}]%
        {orekondy2019prediction}
\bibfield{author}{\bibinfo{person}{Tribhuvanesh Orekondy},
  \bibinfo{person}{Bernt Schiele}, {and} \bibinfo{person}{Mario Fritz}.}
  \bibinfo{year}{2019}\natexlab{b}.
\newblock \showarticletitle{Prediction poisoning: Utility-constrained defenses
  against model stealing attacks}.
\newblock \bibinfo{journal}{\emph{arXiv preprint arXiv:1906.10908}}
  (\bibinfo{year}{2019}).
\newblock


\bibitem[\protect\citeauthoryear{Papernot, McDaniel, Jha, Fredrikson, Celik,
  and Swami}{Papernot et~al\mbox{.}}{2016}]%
        {papernot2016limitations}
\bibfield{author}{\bibinfo{person}{Nicolas Papernot}, \bibinfo{person}{Patrick
  McDaniel}, \bibinfo{person}{Somesh Jha}, \bibinfo{person}{Matt Fredrikson},
  \bibinfo{person}{Z~Berkay Celik}, {and} \bibinfo{person}{Ananthram Swami}.}
  \bibinfo{year}{2016}\natexlab{}.
\newblock \showarticletitle{The limitations of deep learning in adversarial
  settings}. In \bibinfo{booktitle}{\emph{IEEE European symposium on security
  and privacy (EuroS\&P)}}. IEEE, \bibinfo{pages}{372--387}.
\newblock


\bibitem[\protect\citeauthoryear{Pei, Cao, Yang, and Jana}{Pei
  et~al\mbox{.}}{2017}]%
        {pei2017deepxplore}
\bibfield{author}{\bibinfo{person}{Kexin Pei}, \bibinfo{person}{Yinzhi Cao},
  \bibinfo{person}{Junfeng Yang}, {and} \bibinfo{person}{Suman Jana}.}
  \bibinfo{year}{2017}\natexlab{}.
\newblock \showarticletitle{DeepXplore: Automated whitebox testing of deep
  learning systems}. In \bibinfo{booktitle}{\emph{26th Symposium on Operating
  Systems Principles}}. \bibinfo{pages}{1--18}.
\newblock


\bibitem[\protect\citeauthoryear{Qiu, Leng, Guo, Chen, Li, Guo, and Zhu}{Qiu
  et~al\mbox{.}}{2019}]%
        {qiu2019adversarial}
\bibfield{author}{\bibinfo{person}{Yuxian Qiu}, \bibinfo{person}{Jingwen Leng},
  \bibinfo{person}{Cong Guo}, \bibinfo{person}{Quan Chen},
  \bibinfo{person}{Chao Li}, \bibinfo{person}{Minyi Guo}, {and}
  \bibinfo{person}{Yuhao Zhu}.} \bibinfo{year}{2019}\natexlab{}.
\newblock \showarticletitle{Adversarial Defense Through Network Profiling Based
  Path Extraction}. In \bibinfo{booktitle}{\emph{IEEE Conference on Computer
  Vision and Pattern Recognition (CVPR)}}. \bibinfo{pages}{4777--4786}.
\newblock


\bibitem[\protect\citeauthoryear{Rauber, Brendel, and Bethge}{Rauber
  et~al\mbox{.}}{2017}]%
        {rauber2017foolbox}
\bibfield{author}{\bibinfo{person}{Jonas Rauber}, \bibinfo{person}{Wieland
  Brendel}, {and} \bibinfo{person}{Matthias Bethge}.}
  \bibinfo{year}{2017}\natexlab{}.
\newblock \showarticletitle{Foolbox: A python toolbox to benchmark the
  robustness of machine learning models}.
\newblock \bibinfo{journal}{\emph{arXiv preprint arXiv:1707.04131}}
  (\bibinfo{year}{2017}).
\newblock


\bibitem[\protect\citeauthoryear{Reith, Schneider, and Tkachenko}{Reith
  et~al\mbox{.}}{2019}]%
        {reith2019efficiently}
\bibfield{author}{\bibinfo{person}{Robert~Nikolai Reith},
  \bibinfo{person}{Thomas Schneider}, {and} \bibinfo{person}{Oleksandr
  Tkachenko}.} \bibinfo{year}{2019}\natexlab{}.
\newblock \showarticletitle{Efficiently Stealing your Machine Learning Models}.
  In \bibinfo{booktitle}{\emph{18th ACM Workshop on Privacy in the Electronic
  Society}}. \bibinfo{pages}{198--210}.
\newblock


\bibitem[\protect\citeauthoryear{Ross and Doshi-Velez}{Ross and
  Doshi-Velez}{2018}]%
        {ross2018improving}
\bibfield{author}{\bibinfo{person}{Andrew~Slavin Ross} {and}
  \bibinfo{person}{Finale Doshi-Velez}.} \bibinfo{year}{2018}\natexlab{}.
\newblock \showarticletitle{Improving the adversarial robustness and
  interpretability of deep neural networks by regularizing their input
  gradients}. In \bibinfo{booktitle}{\emph{Thirty-second AAAI conference on
  artificial intelligence}}.
\newblock


\bibitem[\protect\citeauthoryear{Shao, Loy, Kang, and Wang}{Shao
  et~al\mbox{.}}{2016}]%
        {shao2016slicing}
\bibfield{author}{\bibinfo{person}{Jing Shao}, \bibinfo{person}{Chen-Change
  Loy}, \bibinfo{person}{Kai Kang}, {and} \bibinfo{person}{Xiaogang Wang}.}
  \bibinfo{year}{2016}\natexlab{}.
\newblock \showarticletitle{Slicing convolutional neural network for crowd
  video understanding}. In \bibinfo{booktitle}{\emph{IEEE Conference on
  Computer Vision and Pattern Recognition}}. \bibinfo{pages}{5620--5628}.
\newblock


\bibitem[\protect\citeauthoryear{Silva}{Silva}{2012}]%
        {silva2012vocabulary}
\bibfield{author}{\bibinfo{person}{Josep Silva}.}
  \bibinfo{year}{2012}\natexlab{}.
\newblock \showarticletitle{A vocabulary of program slicing-based techniques}.
\newblock \bibinfo{journal}{\emph{ACM Computing Surveys (CSUR)}}
  \bibinfo{volume}{44}, \bibinfo{number}{3} (\bibinfo{year}{2012}),
  \bibinfo{pages}{1--41}.
\newblock


\bibitem[\protect\citeauthoryear{Szegedy, Liu, Jia, Sermanet, Reed, Anguelov,
  Erhan, Vanhoucke, and Rabinovich}{Szegedy et~al\mbox{.}}{2015}]%
        {szegedy2015going}
\bibfield{author}{\bibinfo{person}{Christian Szegedy}, \bibinfo{person}{Wei
  Liu}, \bibinfo{person}{Yangqing Jia}, \bibinfo{person}{Pierre Sermanet},
  \bibinfo{person}{Scott Reed}, \bibinfo{person}{Dragomir Anguelov},
  \bibinfo{person}{Dumitru Erhan}, \bibinfo{person}{Vincent Vanhoucke}, {and}
  \bibinfo{person}{Andrew Rabinovich}.} \bibinfo{year}{2015}\natexlab{}.
\newblock \showarticletitle{Going deeper with convolutions}. In
  \bibinfo{booktitle}{\emph{IEEE Conference on Computer Vision and Pattern
  Recognition (CVPR)}}. \bibinfo{pages}{1--9}.
\newblock


\bibitem[\protect\citeauthoryear{Szegedy, Zaremba, Sutskever, Bruna, Erhan,
  Goodfellow, and Fergus}{Szegedy et~al\mbox{.}}{2013}]%
        {szegedy2013intriguing}
\bibfield{author}{\bibinfo{person}{Christian Szegedy},
  \bibinfo{person}{Wojciech Zaremba}, \bibinfo{person}{Ilya Sutskever},
  \bibinfo{person}{Joan Bruna}, \bibinfo{person}{Dumitru Erhan},
  \bibinfo{person}{Ian Goodfellow}, {and} \bibinfo{person}{Rob Fergus}.}
  \bibinfo{year}{2013}\natexlab{}.
\newblock \showarticletitle{Intriguing properties of neural networks}.
\newblock \bibinfo{journal}{\emph{arXiv preprint arXiv:1312.6199}}
  (\bibinfo{year}{2013}).
\newblock


\bibitem[\protect\citeauthoryear{Tan and Le}{Tan and Le}{2019}]%
        {tan2019efficientnet}
\bibfield{author}{\bibinfo{person}{Mingxing Tan} {and} \bibinfo{person}{Quoc~V
  Le}.} \bibinfo{year}{2019}\natexlab{}.
\newblock \showarticletitle{Efficientnet: Rethinking model scaling for
  convolutional neural networks}.
\newblock \bibinfo{journal}{\emph{arXiv preprint arXiv:1905.11946}}
  (\bibinfo{year}{2019}).
\newblock


\bibitem[\protect\citeauthoryear{Tian, Pei, Jana, and Ray}{Tian
  et~al\mbox{.}}{2018}]%
        {tian2018deeptest}
\bibfield{author}{\bibinfo{person}{Yuchi Tian}, \bibinfo{person}{Kexin Pei},
  \bibinfo{person}{Suman Jana}, {and} \bibinfo{person}{Baishakhi Ray}.}
  \bibinfo{year}{2018}\natexlab{}.
\newblock \showarticletitle{DeepTest: Automated testing of
  deep-neural-network-driven autonomous cars}. In
  \bibinfo{booktitle}{\emph{40th International Conference on Software
  Engineering (ICSE)}}. \bibinfo{pages}{303--314}.
\newblock


\bibitem[\protect\citeauthoryear{Tip}{Tip}{1994}]%
        {tip1994survey}
\bibfield{author}{\bibinfo{person}{Frank Tip}.}
  \bibinfo{year}{1994}\natexlab{}.
\newblock \bibinfo{booktitle}{\emph{A survey of program slicing techniques}}.
\newblock \bibinfo{publisher}{Centrum voor Wiskunde en Informatica Amsterdam}.
\newblock


\bibitem[\protect\citeauthoryear{Tramer and Boneh}{Tramer and Boneh}{2018}]%
        {tramer2018slalom}
\bibfield{author}{\bibinfo{person}{Florian Tramer} {and} \bibinfo{person}{Dan
  Boneh}.} \bibinfo{year}{2018}\natexlab{}.
\newblock \showarticletitle{Slalom: Fast, verifiable and private execution of
  neural networks in trusted hardware}.
\newblock \bibinfo{journal}{\emph{arXiv preprint arXiv:1806.03287}}
  (\bibinfo{year}{2018}).
\newblock


\bibitem[\protect\citeauthoryear{Uzuncaova and Khurshid}{Uzuncaova and
  Khurshid}{2007}]%
        {uzuncaova2007kato}
\bibfield{author}{\bibinfo{person}{Engin Uzuncaova} {and}
  \bibinfo{person}{Sarfraz Khurshid}.} \bibinfo{year}{2007}\natexlab{}.
\newblock \showarticletitle{Kato: A program slicing tool for declarative
  specifications}. In \bibinfo{booktitle}{\emph{29th International Conference
  on Software Engineering (ICSE)}}. IEEE, \bibinfo{pages}{767--770}.
\newblock


\bibitem[\protect\citeauthoryear{Wang, Dong, Sun, Wang, and Zhang}{Wang
  et~al\mbox{.}}{2019}]%
        {wang2019model_mutation}
\bibfield{author}{\bibinfo{person}{Jingyi Wang}, \bibinfo{person}{Guoliang
  Dong}, \bibinfo{person}{Jun Sun}, \bibinfo{person}{Xinyu Wang}, {and}
  \bibinfo{person}{Peixin Zhang}.} \bibinfo{year}{2019}\natexlab{}.
\newblock \showarticletitle{Adversarial sample detection for deep neural
  network through model mutation testing}. In \bibinfo{booktitle}{\emph{41st
  International Conference on Software Engineering (ICSE)}}.
  \bibinfo{pages}{1245--1256}.
\newblock


\bibitem[\protect\citeauthoryear{Wang, Pei, Whitehouse, Yang, and Jana}{Wang
  et~al\mbox{.}}{2018a}]%
        {wang2018neurify}
\bibfield{author}{\bibinfo{person}{Shiqi Wang}, \bibinfo{person}{Kexin Pei},
  \bibinfo{person}{Justin Whitehouse}, \bibinfo{person}{Junfeng Yang}, {and}
  \bibinfo{person}{Suman Jana}.} \bibinfo{year}{2018}\natexlab{a}.
\newblock \showarticletitle{Efficient formal safety analysis of neural
  networks}. In \bibinfo{booktitle}{\emph{Advances in Neural Information
  Processing Systems (NeurIPS)}}. \bibinfo{pages}{6367--6377}.
\newblock


\bibitem[\protect\citeauthoryear{Wang, Pei, Whitehouse, Yang, and Jana}{Wang
  et~al\mbox{.}}{2018b}]%
        {wang2018formal}
\bibfield{author}{\bibinfo{person}{Shiqi Wang}, \bibinfo{person}{Kexin Pei},
  \bibinfo{person}{Justin Whitehouse}, \bibinfo{person}{Junfeng Yang}, {and}
  \bibinfo{person}{Suman Jana}.} \bibinfo{year}{2018}\natexlab{b}.
\newblock \showarticletitle{Formal security analysis of neural networks using
  symbolic intervals}. In \bibinfo{booktitle}{\emph{27th {USENIX} Security
  Symposium ({USENIX} Security 18)}}. \bibinfo{pages}{1599--1614}.
\newblock


\bibitem[\protect\citeauthoryear{Wang, Su, Zhang, and Hu}{Wang
  et~al\mbox{.}}{2018c}]%
        {wang2018interpret}
\bibfield{author}{\bibinfo{person}{Yulong Wang}, \bibinfo{person}{Hang Su},
  \bibinfo{person}{Bo Zhang}, {and} \bibinfo{person}{Xiaolin Hu}.}
  \bibinfo{year}{2018}\natexlab{c}.
\newblock \showarticletitle{Interpret neural networks by identifying critical
  data routing paths}. In \bibinfo{booktitle}{\emph{IEEE Conference on Computer
  Vision and Pattern Recognition}}. \bibinfo{pages}{8906--8914}.
\newblock


\bibitem[\protect\citeauthoryear{Weiser}{Weiser}{1981}]%
        {weiser1981slicing}
\bibfield{author}{\bibinfo{person}{Mark Weiser}.}
  \bibinfo{year}{1981}\natexlab{}.
\newblock \showarticletitle{Program slicing}. In
  \bibinfo{booktitle}{\emph{Proceedings of the 5th International Conference on
  Software Engineering}}. IEEE Press, \bibinfo{pages}{439--449}.
\newblock


\bibitem[\protect\citeauthoryear{Xie, Ma, Juefei-Xu, Xue, Chen, Liu, Zhao, Li,
  Yin, and See}{Xie et~al\mbox{.}}{2019}]%
        {xie2019deephunter}
\bibfield{author}{\bibinfo{person}{Xiaofei Xie}, \bibinfo{person}{Lei Ma},
  \bibinfo{person}{Felix Juefei-Xu}, \bibinfo{person}{Minhui Xue},
  \bibinfo{person}{Hongxu Chen}, \bibinfo{person}{Yang Liu},
  \bibinfo{person}{Jianjun Zhao}, \bibinfo{person}{Bo Li},
  \bibinfo{person}{Jianxiong Yin}, {and} \bibinfo{person}{Simon See}.}
  \bibinfo{year}{2019}\natexlab{}.
\newblock \showarticletitle{DeepHunter: a coverage-guided fuzz testing
  framework for deep neural networks}. In \bibinfo{booktitle}{\emph{28th ACM
  SIGSOFT International Symposium on Software Testing and Analysis}}.
  \bibinfo{pages}{146--157}.
\newblock


\bibitem[\protect\citeauthoryear{Yampolskiy}{Yampolskiy}{2011}]%
        {yampolskiy2011ai}
\bibfield{author}{\bibinfo{person}{Roman~V Yampolskiy}.}
  \bibinfo{year}{2011}\natexlab{}.
\newblock \showarticletitle{AI-complete CAPTCHAs as zero knowledge proofs of
  access to an artificially intelligent system}.
\newblock \bibinfo{journal}{\emph{ISRN Artificial Intelligence}}
  \bibinfo{volume}{2012} (\bibinfo{year}{2011}).
\newblock


\bibitem[\protect\citeauthoryear{Zhang, Zhang, Lu, Zhu, and Dong}{Zhang
  et~al\mbox{.}}{2019b}]%
        {zhang2019you}
\bibfield{author}{\bibinfo{person}{Dinghuai Zhang}, \bibinfo{person}{Tianyuan
  Zhang}, \bibinfo{person}{Yiping Lu}, \bibinfo{person}{Zhanxing Zhu}, {and}
  \bibinfo{person}{Bin Dong}.} \bibinfo{year}{2019}\natexlab{b}.
\newblock \showarticletitle{You only propagate once: Accelerating adversarial
  training via maximal principle}. In \bibinfo{booktitle}{\emph{Advances in
  Neural Information Processing Systems}}. \bibinfo{pages}{227--238}.
\newblock


\bibitem[\protect\citeauthoryear{Zhang, Harman, Ma, and Liu}{Zhang
  et~al\mbox{.}}{2020}]%
        {zhang2020mltesting}
\bibfield{author}{\bibinfo{person}{Jie~M Zhang}, \bibinfo{person}{Mark Harman},
  \bibinfo{person}{Lei Ma}, {and} \bibinfo{person}{Yang Liu}.}
  \bibinfo{year}{2020}\natexlab{}.
\newblock \showarticletitle{Machine learning testing: Survey, landscapes and
  horizons}.
\newblock \bibinfo{journal}{\emph{IEEE Transactions on Software Engineering}}
  (\bibinfo{year}{2020}).
\newblock


\bibitem[\protect\citeauthoryear{Zhang, Wang, Xin, and Wu}{Zhang
  et~al\mbox{.}}{2019a}]%
        {zhang2019cheetah}
\bibfield{author}{\bibinfo{person}{Qiao Zhang}, \bibinfo{person}{Cong Wang},
  \bibinfo{person}{Chunsheng Xin}, {and} \bibinfo{person}{Hongyi Wu}.}
  \bibinfo{year}{2019}\natexlab{a}.
\newblock \showarticletitle{CHEETAH: An Ultra-Fast, Approximation-Free, and
  Privacy-Preserved Neural Network Framework based on Joint Obscure Linear and
  Nonlinear Computations}.
\newblock \bibinfo{journal}{\emph{arXiv preprint arXiv:1911.05184}}
  (\bibinfo{year}{2019}).
\newblock


\bibitem[\protect\citeauthoryear{Zhang and Zhu}{Zhang and Zhu}{2018}]%
        {zhang2018visual}
\bibfield{author}{\bibinfo{person}{Quan-shi Zhang} {and}
  \bibinfo{person}{Song-Chun Zhu}.} \bibinfo{year}{2018}\natexlab{}.
\newblock \showarticletitle{Visual interpretability for deep learning: a
  survey}.
\newblock \bibinfo{journal}{\emph{Frontiers of Information Technology \&
  Electronic Engineering}} \bibinfo{volume}{19}, \bibinfo{number}{1}
  (\bibinfo{year}{2018}), \bibinfo{pages}{27--39}.
\newblock


\bibitem[\protect\citeauthoryear{Zhang, Gupta, and Zhang}{Zhang
  et~al\mbox{.}}{2003}]%
        {zhang2003precise}
\bibfield{author}{\bibinfo{person}{Xiangyu Zhang}, \bibinfo{person}{Rajiv
  Gupta}, {and} \bibinfo{person}{Youtao Zhang}.}
  \bibinfo{year}{2003}\natexlab{}.
\newblock \showarticletitle{Precise dynamic slicing algorithms}. In
  \bibinfo{booktitle}{\emph{25th International Conference on Software
  Engineering (ICSE)}}. \bibinfo{pages}{319--329}.
\newblock


\end{thebibliography}

\end{document}